\input amstex
\documentstyle{amsppt}
\pageno=1

\def\Jac{\operatorname {Jac}}
\def\Pch{\operatorname {Pch}}
\def\diag{\operatorname {diag}}
%\document
\NoBlackBoxes
 \topmatter

to appear in {\it Communications in Mathematical Physics}

\

\title
Systems of Hess-Appel'rot type
\endtitle
\author
Vladimir Dragovi\' c{\footnote{vladad\@mi.sanu.ac.yu, Mathematical
Institute, Belgrade, Serbia and Montenegro}} and Borislav Gaji\'
c{\footnote{gajab\@mi.sanu.ac.yu, Mathematical Institute,
Belgrade, Serbia and Montenegro}}
\endauthor
\rightheadtext{Systems of Hess-Appel'rot type}
\leftheadtext{Vladimir Dragovi\'c and Borislav Gaji\'c} \abstract
We construct higher-dimensional generalizations of the classical
Hess-Appel'rot rigid body system. We give a Lax pair with a
spectral parameter leading to an algebro-geometric integration of
this new class of systems, which is closely related to the
integration of the Lagrange bitop performed by us recently and
uses Mumford relation for theta divisors of double unramified
coverings. Based on the basic properties satisfied by such a class
of systems related to bi-Poisson structure, quasi-homogeneity, and
conditions on the Kowalevski exponents, we suggest an axiomatic
approach leading to what we call the "class of systems of
Hess-Appel'rot type".
\endabstract
\endtopmatter

\

\centerline{\bf Contents}

\medskip

\item {1.} Introduction. Starting from the Kowalevski analysis.

\item {2.} Classical Hess-Appel'rot system.

\item {3.} Four-dimensional Hess-Appel'rot system.

\item {4.} The $n$-dimensional Hess-Appel'rot systems.

\item {5.} The decomposition of $so(4)=so(3)\oplus so(3)$ and
integration of the four-di\-men\-si\-o\-nal Hess-Appel'rot system.

\item {6.} Algebro-geometric integration.

\item {7.} A Prym variety.

\item {8.} Isoholomorphicity condition, Mumford's relation and
solutions for $v^k_{LB}$.

\item {9.} The restrictively integrable part -- equations for the
functions $F_i, i=1,...,4$.

\item {10.} Restrictive integrability in an abstract Poisson
algebra settings. Bihamiltonian structures for the Lagrange bitop
and $n$-dimensional Lagrange top.

\item {11.} Back to the Kowalevski properties.

\item {12.} Description of three-dimensional systems of
Hess-Appel'rot type.

\item {} Acknowledgment.

\item {} References.

\newpage

\

\centerline{\bf 1. Introduction. Starting from the Kowalevski
analysis}

\

It is well known that Kowalevski, in her celebrated 1889 paper
[28], starting with a careful analysis of the solutions of the
Euler and the Lagrange case of rigid-body motion, formulated a
problem {\it of describing the parameters $(A, B, C, x_0, y_0,
z_0)$, for which the Euler -- Poisson equations have a general
solution in a form of uniform functions only with moving poles as
singularities.} Here, $I=\diag (A, B, C)$ represents the inertia
operator, and $\chi=(x_0, y_0, z_0)$ is the centre of mass of the
rigid body.

Then, in \S 1 of [28], some necessary  conditions were formulated
and a new case was discovered, now known as Kowalevski case, as a
unique possible beside the cases of Euler and Lagrange. However,
considering the situation where all momenta of inertia are
different, Kowalevski came to a relation  analogue to the
following (see [24]):
$$
x_0\sqrt{A(B-C)}+y_0\sqrt{B(C-A)}+z_0\sqrt{C(A-B)}=0,
$$
and concluded that $x_0=y_0=z_0$, giving the Euler case.

But, it was Appel'rot who noticed in the beginning of 1890's, that
the last relation admits one more case, not mentioned by
Kowalevski:
$$
x_0\sqrt{A(B-C)}+z_0\sqrt{C(A-B)}=0,\quad y_0=0,
$$
under the assumption $A>B>C$. Such systems were considered also by
Hess, even before Appel'rot, in 1890. Such intriguing position
corresponding to the overlook in the Kowalevski paper, made the
Hess-Appel'rot systems very attractive for leading Russian
mathematicians from the end of XIX century. After a few years,
Nekrasov and Lyapunov managed to provide new arguments and they
demonstrated that the Hess-Appel'rot systems didn't satisfy the
condition investigated by Kowalevski, which means that conclusion
of \S 1 of [28] {\it was correct}.

And, from that moment, the Hess-Appel'rot systems were basically
left aside, even in modern times, when new methods of inverse
problems, Lax representations,  finite-zone integrations were
applied to almost all known classical systems, until very
recently.

A few years ago, we  constructed a Lax representation for the
Hess-Appel'rot system (see [15]).

Now, in this paper the first higher - dimensional generalizations
of the Hess-Appel'rot systems are constructed. For each dimension
$n>3$, we give a family of such generalizations. We provide Lax
representations for all new systems, generalizing the Lax pair
from [15]. We show that the new systems are {\it isoholomorphic}.
This class of systems was introduced and
studied in [16], in connection with the Lagrange bitop.

Lax matrices of isoholomorphic systems have specific distributions
of zero entries. Therefore standard integration techniques of
[17], [1] cannot be applied directly. Its integration  requires
more detailed analysis of geometry of Prym varieties and it is
based on Mumford's relation on theta - divisors of unramified
double coverings.

In the present paper, in addition, we perform in detail the
integration procedure in the first higher-dimensional case $n=4$
of new Hess-Appel'rot type systems.

The $L$-operator, a quadratic polynomial in $\lambda$ of the form
$\lambda ^2C+\lambda M + \Gamma, $ in the case $n=4$, satisfies
the condition
$$
L_{12}=L_{21}=L_{34}=L_{43}=0.
$$
Such situation, explicitly excluded by Adler-van Moerbeke (see
[1], Theorem 1) and implicitly by Dubrovin (see [17], Lemma 5 and
Corollary) was studied for the first time in [16]. (A nice and
natural cohomological interpretation of polynomial Lax equations
has been studied in [26].)

Study of the spectral curve and the Baker-Akhiezer function for
the four-di\-men\-si\-o\-nal Hess-Appel'rot systems shows that,
similarly to [16], the dynamics of the system is related to a
Prym variety $\Pi$. It is connected to the evolution of divisors
of certain meromorphic differentials $\Omega^i_j$. From the
condition on zeroes of the Lax matrix, it follows that
differentials
$$
\Omega^1_2,\;\Omega^2_1,\;\Omega^3_4,\;\Omega^4_3
$$
are {\it holomorphic} during the whole evolution. Compatibility of this
requirement with dynamics is based on Mumford's relation (see
 [35], [16])
$$
\Pi^-\subset \Theta ,
$$
where $\Pi^-$ is a translation of a Prym variety $\Pi$, and $\Theta$ is
the theta divisor.

The paper is organized as follows. In Section 2, the definition of
the classical Hess-Appel'rot system  is given and a few of its
basic properties are listed such as the $L-A$ pair from [15] and
the Zhukovskii geometric interpretation from [48, 31]. A
construction of four-dimensional generalizations of the
Hess-Appel'rot system is done in Section 3. In the same Section, a
Lax representation is presented and the spectral curve calculated.
The next Section contains generalizations of the Hess-Appel'rot
systems to all dimensions higher than 4. In the cases $n>4$ not
only invariant relations exist, which are typical for the
Hess-Appel'rot systems, but also values of some of the first
integrals have to be fixed (and equal to zero). Thus, the systems
we construct in the case $n>4$ are also certain generalizations of
the Goryachev-Chaplygin systems (see, for example, [24] for the
definition). The Lax pairs are given in this section as well. In
Section 5, a transformation of coordinates is performed for the
four-dimensional Hess-Appel'rot systems, based on the
decomposition
 $so(4)=so(3)\oplus so(3)$. In
this  manner, the integration of the four-dimensional
Hess-Appel'rot systems reduces to integration of two coupled
three-dimensional systems of Hess-Appel'rot type. Starting with
Section 6, the algebro-geometric integration is performed. The
principal observation is the relationship between the
Baker-Akhiezer functions of the four-dimensional Hess-Appel'rot
system and the Lagrange bitop. Then, in Sections 7 and 8, some
most important facts from the algebro-geometric integration of the
Lagrange bitop, done in [16], are reviewed. Analysis of a Prym
variety $\Pi$ is done and through the Mumford-Dalalyan theory, a
connection between the algebro-geometric and the classical approach from
Section 5 is explained. Differentials $\Omega^i_j$ are defined and
the holomorphicity condition is derived. Therefore, the whole
class of such systems is called {\bf isoholomorphic} systems. The
crucial point is application of {\it Mumford's relation on
theta-divisors of unramified double coverings} to derive formulae
in theta-functions for such systems. Their dynamics is realized on
the odd part of the generalized Jacobian, which is obtained by
gluing of the infinite points of the spectral curve. In Section 9,
additional equations, which differ the cases of the Lagrange bitop
and higher-dimensional Hess-Appel'rot systems are derived. In the
final part of the paper, the characteristic properties, common for
the Hess-Appel'rot system and its higher-dimensional
generalizations are studied. The most relevant ones are abstracted
as the axioms of the {\bf class of systems of Hess-Appel'rot
type}. In this way, in Section 10, after analysis of relevant
Poisson structures, the {\it Hamilton perturbation} and {\it
bi-Poisson } axioms are formulated. These axioms give very simple
and geometrically transparent description of the systems of
Hess-Appel'rot type. Namely, suppose bi-Poisson structure
$\{\cdot,\cdot\}_1+\lambda\{\cdot,\cdot\}_2$ is given, with a
bihamiltonian system with the Hamiltonian $H_0$ corresponding to
the first structure. Further, let $f_1,\dots, f_k$ be the
commuting integrals of the system $(H_0,\{\cdot,\cdot\}_1)$, which
are Casimirs for the second structure $\{\cdot,\cdot\}_2$. Then,
the systems of Hess-Appel'rot type are Hamiltonian with respect to
the first structure with a Hamiltonian
$$
H=H_0+\sum_{l=1}^k J_lb_lf_l,
$$
where $J_l$ are constants and $b_l$ are certain functions on the phase
space. The invariant relations are
$$
f_l=0,\quad l=1,\dots, k.
$$
Thus, the invariant manifolds are symplectic leaves of the second
structure.

In Section 11, a Kowalevski analysis is
performed. As a result, the {\it quasi-homogeneity} and the {\it
arithmetic} axiom
 are formulated, providing characterization
of Hess-Appel'rot systems in terms of arithmetic conditions on
Kowalevski exponents. This gives strong constraints on the
functions $b_l$ in the above expressions.

In this way, the study of Hess-Appel'rot systems, in a sense,
reaches its historical origins of Kowalevski, Appel'rot, Lyapunov
and others, as briefly mentioned above.

Finally, based on these axioms we study three-dimensional
Hess-Appel'rot systems and formulate conditions which determine
uniquely the classical Hess-Appel'rot system among them. This
confirms the reasonability of the chosen axioms. Classification of
higher-dimensional Hess-Appel'rot systems looks like an
interesting problem. We hope that detailed analysis of
dynamical properties of systems of Hess-Appel'rot type will
deserve sufficient attention.

\

\centerline{\bf 2. Classical Hess-Appel'rot system.}

\

The Euler-Poisson equations of the motion of a heavy rigid body in the moving frame
are [24]:
$$
\aligned
\dot {\bold M}&={\bold M}\times {\bold \Omega}+
{\bold \Gamma}\times{\bold \chi},\\
\dot {\bold \Gamma}&={\bold \Gamma}\times{\bold \Omega}\\
{\bold \Omega}&=\tilde{J}{\bold M},\ \ \tilde{J}=\diag(\tilde{J}_1, \tilde{J}_2, \tilde{J}_3),
\endaligned
\tag1
$$
where $\bold M$ is the kinetic momentum vector, $\bold \Omega$ the
angular velocity, $\tilde{J}$ a diagonal matrix, the inverse of inertia
operator, $\bold \Gamma$ a unit vector fixed in the space and
$\bold \chi$ is the radius vector of the centre of masses.

It is well known ([24]) that equations (1) have three integrals of
motion:
$$
\aligned
F_1&=\frac 12\langle {\bold M}, {\bold\Omega}\rangle+\langle{\bold\Gamma}, {\bold\chi}\rangle\\
F_2&=\langle {\bold M}, {\bold\Gamma}\rangle\\
F_3&=\langle {\bold \Gamma}, {\bold\Gamma}\rangle=1.
\endaligned
\tag2
$$

Thus, for complete integrability, one integral more is necessary
[24]. Let $\tilde{J}_1<\tilde{J}_2<\tilde{J}_3$ and ${\bold \chi}=(x_0, y_0, z_0)$. Hess
in [27] and Appel'rot in [4] found that if the inertia momenta and
the radius vector of the centre of masses satisfy the conditions
$$
\aligned
y_0&=0\\
x_0\sqrt{\tilde{J}_3-\tilde{J}_2}&+z_0\sqrt{\tilde{J}_2-\tilde{J}_1}=0,
\endaligned
\tag3
$$
then the surface
$$
F_4=M_1x_0+M_3z_0=0
\tag4
$$
is invariant. Integration of such a system by classical techniques
can be found in [24]. In [15], an L-A pair for the Hess-Appel'rot
system is constructed:
$$
\aligned
\dot L(\lambda)&=[L(\lambda), A(\lambda)],\\
L(\lambda)=\lambda^2 C+ \lambda M+\Gamma,\ \
A(\lambda)&=\lambda\chi +\Omega, \ \ C=\frac{1}{\tilde{J}_2}\chi,
\endaligned
$$
where skew-symmetric matrices represent vectors denoted by the
same letter.
 Also, basic steps in algebro-geometric integration
procedure are given in [15].

{\it The Zhukovskii  geometric interpretation of the conditions}
(3) [48, 31]. Let us consider the ellipsoid
$$
\frac {M_1^2}{\tilde{J}_1}+\frac {M_2^2}{\tilde{J}_2}+\frac {M_3^2}{\tilde{J}_3}=1,
$$
and the plane containing the middle axis and  intersecting the
ellipsoid at a circle. Denote by $l$ the normal to the plane,
which passes through the fixed point $O$. Then the condition (3)
means that the centre of masses  lies on the line $l$.

Having this interpretation in mind, we choose a basis of moving
frame  such that the third axis is $l$, the second one is directed
along the middle axis of the ellipsoid, and the first one is
chosen according to the orientation of the orthogonal frame. In
this basis (see [13]), the particular integral (4) becomes
$$
F_4=M_3=0,
$$
matrix $\tilde{J}$ obtains the form:
$$
J=\left(\matrix J_1&0&J_{13}\\
                0&J_1&0\\
                J_{13}&0&J_3
        \endmatrix\right),
$$
and ${\bold \chi}=(0, 0, z_0)$.  This will serve us as a
motivation for a definition of the four-dimensional Hess-Appel'rot
system.

\

\

\centerline{\bf 3. Four-dimensional Hess-Appel'rot system.}

\

The Euler-Poisson equations of motion of a heavy rigid body fixed
at a point are Hamiltonian on the Lie algebra $e(3)$, which is the
semi-direct product of Lie algebras $R^3$ and $so(3)$. Since $R^3$
is isomorphic to $so(3)$,  there are two natural
higher-dimensional generalizations of Euler-Poisson equations. One
is to Lie algebra $e(n)=R^n\times so(n)$, and the second one,
given by Ratiu in [37], is to the semi-direct product $so(n)\times
so(n)$. The main result of this Section is a construction of an
analogue of the Hess-Appel'rot system on $so(n)\times so(n)$.

Equations of a heavy $n$-dimensional rigid body on $so(n)\times
so(n)$, introduced by Ratiu in [37], are:
$$
\aligned
\dot M&=[M, \Omega]+[\Gamma, \chi]\\
\dot\Gamma&=[\Gamma,\Omega],
\endaligned
\tag5
$$
where $M, \Omega, \Gamma,\chi\in so(n)$, and $\chi$ is a  constant
matrix. We will suppose that
$$
\Omega=JM+MJ,
$$
where $J$ is a constant symmetric matrix.  First, in this section,
we consider equations (5) in dimension four. Motivated by the
Zhukovskii geometric interpretation given at the end of the
previous section, we start  with the following definition

\proclaim{Definition 1} The four-dimensional Hess-Appel'rot system
is
 described by the equations {\rm(5)} and satisfies the conditions:

\item{{\rm a)}}
$$\Omega=MJ+JM,\quad J=\left (\matrix J_1&0&J_{13}&0\\
                    0&J_1&0&J_{24}\\
                    J_{13}&0&J_3&0\\
                    0&J_{24}&0&J_3
\endmatrix\right)
\tag6
$$
\item {{\rm b)}}
$$
\chi=\left(\matrix 0&\chi_{12}&0&0\\
                    -\chi_{12}&0&0&0\\
                    0&0&0&\chi_{34}\\
                    0&0&-\chi_{34}&0
                    \endmatrix\right).
$$
\endproclaim

The invariant surfaces are determined in the next lemma.

\proclaim{Lemma 1} For the four-dimensional Hess-Appel'rot system,
the following relations take place:
$$
\aligned
\dot M_{12}&=J_{13}(M_{13}M_{12}+M_{24}M_{34})+J_{24}(M_{13}M_{34}+M_{12}M_{24}),\\
\dot M_{34}&=J_{13}(-M_{13}M_{34}-M_{12}M_{24})+J_{24}(-M_{13}M_{12}-M_{24}M_{34}).
\endaligned
$$
In particular, if $M_{12}=M_{34}=0$ hold at the initial moment,
then the same relations are  satisfied during the  evolution in
time.
\endproclaim

Proof follows by direct calculations from equations (5), using (6).

Thus, in the four-dimensional Hess-Appel'rot case, there are two
invariant relations
$$
M_{12}=0,\quad M_{34}=0.
\tag{7}
$$

Now we will give another definition of the four-dimensional
Hess-Appel'rot conditions, starting from a basis where the matrix
$J$ is diagonal in.

Let $\tilde J=\diag(\tilde J_1,
\tilde J_2, \tilde J_3, \tilde J_4)$.

\proclaim{Definition 1'} The  four-dimensional Hess-Appel'rot
system is
 described by the equations (5) and satisfies the conditions:

\item {{\rm a)}}
$$
\Omega=M\tilde J+\tilde J M,\ \ \tilde J=\diag(\tilde J_1, \tilde J_2,
\tilde J_3, \tilde J_4),
$$
\item {{\rm b)}}
$$
\tilde\chi=\left(\matrix 0&\tilde\chi_{12}&0&\tilde\chi_{14}\\
                    -\tilde\chi_{12}&0&\tilde\chi_{23}&0\\
                    0&-\tilde\chi_{23}&0&\tilde\chi_{34}\\
                    -\tilde\chi_{14}&0&-\tilde\chi_{34}&0
                    \endmatrix\right),
$$
\item {c)}
$$
\aligned
\tilde J_3-\tilde J_4 &=\tilde J_2-\tilde J_1,\\
\frac{\tilde{J_3}-\tilde{J_1}}{\sqrt{1+t_1^2}}&=
\frac{\tilde{J_4}-\tilde{J_2}}{\sqrt{1+t_2^2}}
\\
\endaligned
$$
where
$$
\aligned
t_1 & :=\frac {2(\tilde \chi_{14}\tilde \chi_{34}-\tilde \chi_{12}\tilde
\chi_{23})}{\tilde \chi_{14}^2-\tilde \chi_{34}^2+\tilde \chi_{12}^2-\tilde
\chi_{23}^2},\\
t_2 & :=\frac {2(\tilde \chi_{14}\tilde \chi_{12}-\tilde
\chi_{23}\tilde \chi_{34})}{-\tilde \chi_{14}^2-\tilde
\chi_{34}^2+\tilde \chi_{12}^2+\tilde
\chi_{23}^2}.\\
\endaligned
$$
\endproclaim

\proclaim{Proposition 1} There exists a bi-correspondence between
sets of data from Definition 1 and Definition 1'.
\endproclaim

\demo{Proof} From $\tilde J=S^TJS$, where
$$
S=\left(\matrix \cos\varphi&0&\sin\varphi&0\\
                0&\cos\varphi_1&0&\sin\varphi_1\\
                -\sin\varphi&0&\cos\varphi&0\\
                0&-\sin\varphi_1&0&\cos\varphi_1
                \endmatrix\right),
$$
and $\varphi=\frac12 \arctan\frac{2J_{13}}{J_3-J_1}$, $\varphi_1=\frac12 \arctan\frac{2J_{24}}{J_3-J_1}$, we have
$$
\tilde J=\diag\left(\frac{J_1+J_3-A}{2},
\frac{J_1+J_3-A_1}{2},\frac{J_1+J_3+A}{2},
\frac{J_1+J_3+A_1}{2}\right).
$$
Here $A=\sqrt{(J_3-J_1)^2+4J_{13}^2},\;$
$A_1=\sqrt{(J_3-J_1)^2+4J_{24}^2}$ . The first part in the
Definition 1'c)
$$
\tilde J_3-\tilde J_4=\tilde J_2-\tilde J_1,
$$
follows from these relations.

From $\tilde\chi=S^T\chi S$, we have
$$
\tilde\chi=S^T\chi S=\left(\matrix 0&\tilde\chi_{12}&0&\tilde\chi_{14}\\
                    -\tilde\chi_{12}&0&\tilde\chi_{23}&0\\
                    0&-\tilde\chi_{23}&0&\tilde\chi_{34}\\
                    -\tilde\chi_{14}&0&-\tilde\chi_{34}&0
                    \endmatrix\right),
$$
where
$$
\aligned
\tilde\chi_{12} &=\chi_{12}\cos\varphi\cos\varphi_1+\chi_{34}\sin\varphi\sin\varphi_1,\\
\tilde\chi_{14} &=\chi_{12}\cos\varphi\sin\varphi_1-\chi_{34}\sin\varphi\cos\varphi_1,\\
\tilde\chi_{23} &=-\chi_{12}\sin\varphi\cos\varphi_1+\chi_{34}\cos\varphi\sin\varphi_1,\\
\tilde\chi_{34} &=\chi_{12}\sin\varphi\sin\varphi_1+\chi_{34}\cos\varphi\cos\varphi_1.\\
\endaligned
$$
From the last formulae, it follows:
$$
(\tilde\chi_{12}\sin\varphi+\tilde\chi_{23}\cos\varphi)\cos\varphi_1+(\tilde\chi_{14}\sin\varphi-\tilde\chi_{34}\cos\varphi)\sin\varphi_1=0,
\tag 8
$$
$$
(\tilde\chi_{12}\cos\varphi-\tilde\chi_{23}\sin\varphi)\sin\varphi_1-(\tilde\chi_{14}\cos\varphi+\tilde\chi_{34}\sin\varphi)\cos\varphi_1=0.
\tag 9
$$

From (8) and (9):
$$
\aligned
\tan 2\varphi & =\frac {2(\tilde \chi_{14}\tilde \chi_{34}-\tilde \chi_{12}\tilde \chi_{23})}{\tilde \chi_{14}^2-\tilde \chi_{34}^2+\tilde \chi_{12}^2-\tilde \chi_{23}^2}=:t_1,\\
\tan 2\varphi_1 & =\frac {2(\tilde \chi_{14}\tilde \chi_{12}-\tilde \chi_{23}\tilde \chi_{34})}{-\tilde \chi_{14}^2-\tilde \chi_{34}^2+\tilde \chi_{12}^2+\tilde \chi_{23}^2}=:t_2.\\
\endaligned
$$
Thus, we get
$$
\aligned (J_1-J_3)^2&=\frac{(\tilde J_3-\tilde J_1)^2}{1+t_1^2},\\
(J_1-J_3)^2&=\frac{(\tilde J_4-\tilde J_2)^2}{1+t_2^2}.\\
\endaligned
$$
From the last formulae, we come to the last part of the Definition
1' c. This finishes the proof. \qed

\enddemo

{\bf Note.} 1) In the case
$$
J_{24}\ne 0, \chi_{34}=0,
$$
there is an additional relation
$$
\tilde \chi_{12}\tilde \chi_{34}+\tilde \chi_{14}\tilde
\chi_{23}=0.
$$
It follows from the system
$$
\aligned
\tilde\chi_{12}\sin\varphi+\tilde\chi_{23}\cos\varphi &=0,\\
\tilde\chi_{14}\sin\varphi-\tilde\chi_{34}\cos\varphi &=0,\\
\endaligned
$$
as a consequence of (8, 9).

2) In the  case
$$
J_{24}= 0, \chi_{34}=0,
$$
additional relations are
$$
\tilde \chi_{34}=\tilde \chi_{14}=0,
$$
and the second relation from the Definition 1' c) can be replaced by
the relation
$$
\tilde\chi_{12}\sqrt{\tilde J_2-\tilde J_1}+
\tilde\chi_{23}\sqrt{\tilde J_3-\tilde J_2}=0.
$$
Notice the similarity of the last condition   with the condition
(3) for the classical three-dimensional case. (By ignoring the last
coordinate one can recover the three-dimensional Hess-Appel'rot case.)

\proclaim {Theorem 1} The four-dimensional Hess-Appel'rot system
 has the following Lax representation
$$
\aligned
\dot L(\lambda)&=[L(\lambda), A(\lambda)],\\
L(\lambda)=\lambda^2 C+ \lambda M+\Gamma,\ \ A(\lambda)&=\lambda\chi +\Omega,
\ \ C=\frac1{J_1+J_3}\chi.
\endaligned
$$
\endproclaim

\demo{Proof} Proof follows from
$$
[C,\Omega]+[M,\chi]=
 \frac1{J_1+J_3}\left(\matrix
0&0&D_{13}&0\\
0&0&0&D_{24}\\
-D_{13}&0&0&0\\
0&-D_{24}&0&0
\endmatrix\right),
$$
where
$$
\aligned
D_{13}&=-\chi_{12}(J_{13}M_{12}+J_{24}M_{34})+\chi_{34}(J_{13}M_{34}+J_{24}M_{12}),\\
D_{24}&=-\chi_{12}(J_{13}M_{34}+J_{24}M_{12})+\chi_{34}(J_{13}M_{12}+J_{24}M_{34}),
\endaligned
$$
using  relations (7).\qed
\enddemo

One can calculate the spectral polynomial for the four-dimensional
Hess-Appel'rot system:
$$
p(\lambda, \mu)=\det(L(\lambda)-\mu\cdot 1)=
\mu^4+P(\lambda)\mu^2+Q(\lambda)^2,
$$
where
$$
\aligned
P(\lambda)&=a\lambda^4+b\lambda^3+c\lambda^2+d\lambda+e\\
Q(\lambda)&=f\lambda^4+g\lambda^3+h\lambda^2+i\lambda+j
\endaligned
$$
$$
\aligned
a&=C_{12}^2+C_{34}^2,\\
b&=2C_{12}M_{12}+2C_{34}M_{34}(=0),\\
c&=M_{13}^2+M_{14}^2+M_{23}^2+M_{24}^2+M_{12}^2+M_{34}^2+2C_{12}\Gamma _{12}+2C_{34}\Gamma_{34},\\
d&=2\Gamma _{12}M_{12}+2\Gamma _{13}M_{13}+2\Gamma _{14}M_{14}+2\Gamma _{23}
M_{23}+2\Gamma _{24}M_{24}+2\Gamma _{34}M_{34}\\
e&=\Gamma _{12}^2+\Gamma _{13}^2+\Gamma _{14}^2+\Gamma _{23}^2+\Gamma _{24}^2
+\Gamma _{34}^2,\\
f&=C_{12}C_{34}\\
g&=C_{12}M_{34}+C_{34}M_{12}(=0),\\
h&=\Gamma _{34}C_{12}+\Gamma_{12}C_{34}+M_{12}M_{34}+M_{23}M_{14}-M_{13}M_{24},\\
i&=M_{34}\Gamma _{12}+M_{12}\Gamma _{34}+M_{14}\Gamma _{23}+M_{23}\Gamma _{14}-
\Gamma _{13}M_{24}-\Gamma _{24}M_{13},\\
j&=\Gamma _{34}\Gamma _{12}+\Gamma _{23}\Gamma _{14}-\Gamma _{13}\Gamma _{24}.
\endaligned
$$
Let us consider standard Poisson structure on semidirect product $so(4)\times so(4)$.
The functions $d,e,i,j$ are Casimir functions (see [37]), $c, h$
are first integrals, and $b=0, g=0$ are the invariant relations.
General orbits of co-adjoint action are eight-dimensional, thus for
complete integrability one needs four independent integrals in involution.

\

\

\centerline{\bf 4. The $n$-dimensional Hess-Appel'rot systems.}

\

In this Section, we introduce Hess-Appel'rot systems of arbitrary
dimension.

\proclaim{Definition 2} The $n$-dimensional Hess-Appel'rot system
is described by the equations {\rm(5)}, and satisfies the
conditions: \item{{\rm a)}}
$$
\Omega=JM+MJ,\ \ J=\left (\matrix J_1&0&J_{13}&0&0&...&0\\
                                0&J_1&0&J_{24}&0&...&0\\
                                J_{13}&0&J_3&0&0&...&0\\
                                0&J_{24}&0&J_3&0&...&0\\
                                0&0&0&0&0&...&0\\
                                .&.&.&.&.&...&.\\
                                .&.&.&.&.&...&.\\
                                0&0&0&0&0&...&J_3
                                \endmatrix\right),
\tag{10}
$$
\item {{\rm b)}}
$$
\chi=\left (\matrix 0&\chi_{12}&0&...&0\\
                    -\chi_{12}&0&0&...&0\\
                    0&0&0&...&0\\
                    0&0&0&...&0\\
                    .&.&.&...&.\\
                    .&.&.&...&.\\
                    0&0&0&...&0
                    \endmatrix\right).
$$
\endproclaim
Direct calculations from (5) using (6) give the  following lemma:

\proclaim{Lemma 2} For the $n$-dimensional Hess-Appel'rot system,
the following relations are satisfied: \item {{\rm a)}}
$$
\aligned
\dot M_{12}&=J_{13}(M_{12}M_{13}+M_{24}M_{34}+\sum_{p=5}^n M_{2p}M_{3p})+\\
&J_{24}(M_{12}M_{24}+M_{13}M_{34}-\sum_{p=5}^nM_{1p}M_{4p})\\
\dot M_{34}&=-J_{13}(M_{13}M_{34}+M_{24}M_{12}+\sum_{p=5}^nM_{1p}
M_{p4})-\\
&J_{24}(M_{13}M_{12}+M_{24}M_{34}+\sum_{p=5}^nM_{2p} M_{3p}),\\
\dot M_{3p}&=-J_{13}(M_{13}M_{3p}+M_{2p}M_{12})-J_{24}(M_{34}M_{2p}+M_{23}M_{4p})+\\
&M_{34}\Omega_{4p}-\Omega_{34}M_{4p}+\sum_{k=5}^n(M_{3k}\Omega_{kp}-\Omega_{3k}M_{4p}),\
p>4,\\
\dot M_{4p}&=J_{13}(-M_{14}M_{3p}+M_{1p}M_{34})+J_{24}(M_{12}M_{1p}-M_{24}M_{4p})-\\
&M_{34}\Omega_{3p}+\Omega_{34}M_{3p}+\sum_{k=5}^n(M_{4k}\Omega_{kp}-\Omega_{4k}M_{4p}),\
p>4,\\
\endaligned
$$
\item {{\rm b)}}
$$
\dot M_{kl}=0,\ \ \ k,l>4.
$$
\item{{\rm c)}}
The $n$-dimensional Hess-Appel'rot case has the following
system of invariant relations
$$
M_{12}=0,\ \ M_{lp}=0,\ \ l,p\ge 3. \tag {11}
$$
\endproclaim
By  diagonalizing the matrix $J$, we come to another  definition

\proclaim{Definition 2'} The $n$-dimensional Hess-Appel'rot system
is described by the equations {\rm(5)}, and satisfies the
conditions \item{{\rm a)}}
$$
\Omega=\tilde J M+M\tilde J,\ \ \tilde J=\diag(\tilde J_1, \tilde
J_2, \tilde J_3, \tilde J_4,...,\tilde J_4),
$$
\item {{\rm b)}}
$$
\tilde\chi=\left (\matrix 0&\tilde\chi_{12}&0&\tilde\chi_{14}&...&0\\
                    -\tilde\chi_{12}&0&\tilde \chi_{23}&0&...&0\\
                    0&-\tilde \chi_{23}&0&\tilde\chi_{34}&...&0\\
                    -\tilde\chi_{14}&0&-\tilde\chi_{34}&0&...&0\\
                    .&.&.&.&...&.\\
                    .&.&.&.&...&.\\
                    0&0&0&0&...&0
                    \endmatrix\right),
$$
\item {{\rm c)}}
$$
\aligned
\tilde J_3-\tilde J_4 &=\tilde J_2-\tilde J_1,\\
\frac{\tilde{J_3}-\tilde{J_1}}{\sqrt{1+t_1^2}}&=
\frac{\tilde{J_4}-\tilde{J_2}}{\sqrt{1+t_2^2}}
\\
\tilde\chi_{12}\tilde\chi_{34}&+\tilde\chi_{14}\tilde\chi_{23}=0
\endaligned
$$
where
$$
\aligned t_1 & :=\frac {2(\tilde \chi_{14}\tilde \chi_{34}-\tilde
\chi_{12}\tilde \chi_{23})}{\tilde \chi_{14}^2-\tilde
\chi_{34}^2+\tilde \chi_{12}^2-\tilde
\chi_{23}^2},\\
t_2 & :=\frac {2(\tilde \chi_{14}\tilde \chi_{12}-\tilde
\chi_{23}\tilde \chi_{34})}{-\tilde \chi_{14}^2-\tilde
\chi_{34}^2+\tilde \chi_{12}^2+\tilde
\chi_{23}^2}.\\
\endaligned
$$

\endproclaim
As in the dimension four, there is an equivalence of the definitions.

\proclaim{Proposition 2} There exists a bi-correspondence between
sets of data from Definition 2 and Definition 2'.
\endproclaim

Proof follows the steps in Proposition 1.

Next theorem gives a Lax pair for the $n$-dimensional
Hess-Appel'rot system.

\proclaim {Theorem 2} The $n$-dimensional Hess-Appel'rot system
has the following Lax pair
$$
\aligned
\dot L(\lambda)&=[L(\lambda), A(\lambda)],\\
L(\lambda)=\lambda^2 C+ \lambda M+\Gamma,\ \ A(\lambda)&=\lambda\chi +\Omega,
\ \ C=\frac1{J_1+J_3}\chi.
\endaligned
$$
\endproclaim
\demo{Proof} The statement follows from
$$
\multline
[C,\Omega]+[M,\chi]=-\frac{\chi_{12}}{J_1+J_3}\cdot\\
\cdot\left(\matrix
0&0&J_{13}M_{12}+J_{24}M_{34}&0&J_{24}M_{45}&...&J_{24}M_{4n}\\
&0&0&J_{13}M_{34}+J_{24}M_{12}&J_{13}M_{45}&...&J_{13}M_{3n}\\
&&0&0&0&...&0\\
&&&...&\\
&&&&...&\\
&&&&&...&\\
&&&&&&0
\endmatrix\right)
\endmultline
$$
and relations (11). \qed
\smallskip
{\bf Note.} Let us note that invariant relations (11) exist in a
more general case, with matrix $J$ given by:
$$
J:=\left( \matrix J_1&0&J_{13}&J_{14}&...&J_{1n}\\
                0&J_1&J_{23}&J_{24}&...&J_{2n}\\
                J_{13}&J_{23}&J_3&0&...&0\\
                J_{14}&J_{24}&0&J_3&...&0\\
                &&&&...&\\
                &&&&...&\\
                &&&&...&\\
                J_{1n}&J_{2n}&0&0&...&J_3
                \endmatrix
                \right)
$$
But, using transformations $J\mapsto T^tJT$, where T is a
block-diagonal matrix with $2\times2$ - block $A \in SO(2)$ and
$(n-2)\times (n-2)$-block $B\in SO(n-2)$  on the diagonal, such a
more general matrix $J$ can be transformed to the case considered
above.

Let us mention again that the Goryachev-Chaplygin system is a
classical case integrable for the fixed level of a first integral.
According to Lema 2b) and 2c) in the $n$-dimensional Hess-Appel'rot
system we also fix values of certain first integrals. But also, we have invariant
relations which do not exist in the Goryachev-Chaplygin case.

\

\

\centerline{\bf 5. The  decomposition  $so(4)=so(3)\oplus so(3)$
and} \centerline {\bf integration of the four-dimensional
Hess-Appel'rot system.}

\

Starting from the well-known decomposition  $so(4)=so(3)\oplus
so(3)$, let us introduce
$$
M_1=\frac 12 (M_++M_-)\qquad M_2=\frac 12 (M_+-M_-),
$$
(and similarly for $\Omega, \Gamma, \chi$), where $M_+, M_-$ are
vectors in $R^3$ defined with following correspondence between two
three-dimensional vectors and four-dimensional antisymmetric
matrices
$$
(M_+,M_-)\rightarrow\pmatrix
0 & -M^3_{+} & M^2_{+} & -M^1_{-}\\
M^3_{+} & 0 & -M^1_{+} & -M^2_{-}\\
-M^2_{+} & M^1_{+} & 0 & -M^3_{-}\\
M^1_{-} & M^2_{-} & M^3_{-} & 0
\endpmatrix .
$$
Then, equations of the motion become
$$
\aligned
\dot M_1&=2(M_1\times\Omega_1+\Gamma_1\times\chi_1)\qquad
\dot \Gamma_1=2(\Gamma_1\times\Omega_1)\\
\dot M_2&=2(M_2\times\Omega_2+\Gamma_2\times\chi_2)\qquad
\dot \Gamma_2=2(\Gamma_2\times\Omega_2),
\endaligned
\tag{12}
$$
and
$$
\chi_1=(0,0,-\frac12 (\chi_{12}+\chi_{34})),\quad
\chi_2=(0,0,-\frac12 (\chi_{12}-\chi_{34})).
$$
Integrals of the motion are
$$
\aligned
\langle M_i, M_i\rangle+2\frac{1}{J_1+J_3}\langle \chi_i, \Gamma_i\rangle&=h_i,\\
\langle\Gamma_i,\Gamma_i\rangle&=1,\qquad i=1,2,\\
\langle M_i,\Gamma_i\rangle&=c_i,\\
\langle\chi_i,M_i\rangle&=0.
\endaligned
\tag{13}
$$
Connections between $M$ and $\Omega$ are
$$
\aligned
\Omega_1&=( (J_1+J_3)M_{(1)1}-(J_{13}-J_{24})M_{(2)3}, (J_1+J_3)M_{(1)2},\\
&(J_1+J_3)M_{(1)3}+(J_1-J_3)M_{(2)3}-(J_{13}+J_{24})M_{(2)1}),\\
\Omega_2&=( (J_1+J_3)M_{(2)1}-(J_{13}+J_{24})M_{(1)3}, (J_1+J_3)M_{(2)2},\\
&(J_1+J_3)M_{(2)3}+(J_1-J_3)M_{(1)3}-(J_{13}-J_{24})M_{(1)1}),
\endaligned
$$
where $M_{(i)j}$ is the $j$-th component of vector $M_i$. Using
these expressions, equations (12) can be rewritten in the
following form:
$$
\aligned
\dot M_{(1)1}=&2[ (J_1-J_3)M_{(1)2}M_{(2)3}-(J_{13}+J_{24})M_{(1)2}M_{(2)1}
+\Gamma_{(1)2}\chi_{(1)3}],\\
\dot M_{(1)2}=&2[-(J_1-J_3)M_{(2)3}M_{(1)1}-(J_{13}-J_{24})M_{(1)3}M_{(2)3}+\\
&(J_{13}+J_{24})M_{(1)1}M_{(2)1}-\Gamma_{(1)1}\chi_{(1)3}],\\
\dot M_{(1)3}=&2(J_{13}-J_{24})M_{(1)2}M_{(2)3},\\
\dot \Gamma_{(1)1}=&2[\Gamma_{(1)2}((J_1+J_3)M_{(1)3}+(J_1-J_3)M_{(2)3}-
(J_{13}+J_{24})M_{(2)1})-\\
&\Gamma_{(1)3}(J_1+J_3)M_{(1)2}],\\
\dot \Gamma_{(1)2}=&2[\Gamma_{(1)3}((J_1+J_3)M_{(1)1}-(J_{13}-J_{24})M_{(2)3})-\\
&\Gamma_{(1)1}((J_1+J_3)M_{(1)3}+(J_1-J_3)M_{(2)3}-(J_{13}+J_{24})M_{(2)1})],\\
\dot \Gamma_{(1)3}=&2[\Gamma_{(1)1}(J_1+J_3)M_{(1)2}-\Gamma_{(1)2}((J_1+J_3)M_{(1)1}-
(J_{13}-J_{24})M_{(2)3})],
\endaligned
\tag{14}
$$
and
$$
\aligned
\dot M_{(2)1}=&2[ (J_1-J_3)M_{(2)2}M_{(1)3}-(J_{13}-J_{24})M_{(2)2}M_{(1)1}
+\Gamma_{(2)2}\chi_{(2)3}],\\
\dot M_{(2)2}=&2[-(J_1-J_3)M_{(1)3}M_{(2)1}-(J_{13}+J_{24})M_{(2)3}M_{(1)3}+\\
&(J_{13}-J_{24})M_{(2)1}M_{(1)1}-\Gamma_{(2)1}\chi_{(2)3}],\\
\dot M_{(2)3}=&2(J_{13}+J_{24})M_{(2)2}M_{(1)3},\\
\dot \Gamma_{(2)1}=&2[\Gamma_{(2)2}((J_1+J_3)M_{(2)3}+(J_1-J_3)M_{(1)3}-
(J_{13}-J_{24})M_{(1)1})-\\
&\Gamma_{(2)3}(J_1+J_3)M_{(2)2}],\\
\dot \Gamma_{(2)2}=&2[\Gamma_{(2)3}((J_1+J_3)M_{(2)1}-(J_{13}+J_{24})M_{(1)3})-\\
&\Gamma_{(2)1}((J_1+J_3)M_{(2)3}+(J_1-J_3)M_{(1)3}-(J_{13}-J_{24})M_{(1)1})],\\
\dot \Gamma_{(2)3}=&2[\Gamma_{(2)1}(J_1+J_3)M_{(2)2}-\Gamma_{(2)2}((J_1+J_3)M_{(2)1}-
(J_{13}+J_{24})M_{(1)3})].
\endaligned
\tag{15}
$$

From the equations (14) and (15), it follows that
$M_{(1)3}=M_{(2)3}=0$, giving two invariant relations introduced
before.

Now, we are going to proceed the  integration in a classical
manner.

First, let us introduce coordinates $K_i$ and $l_i$ as follows:
$$
M_{(i)1}=K_i\sin l_i,\qquad M_{(i)2}=K_i\cos l_i,\qquad i=1,2.
$$
From the sixth equation of (14), using integrals (13), we have
that
$$
\dot\Gamma_{(1)3}^2=4(J_1+J_3)^2\left[(1-\Gamma_{(1)3}^2)
(h_1-\frac2{J_1+J_3}\chi_{(1)3}\Gamma_{(1)3})-c_1^2\right]=
P_3(\Gamma_{(1)3}).
$$
Thus $\Gamma_{(1)3}$ can be solved by an elliptic quadrature. Also
from the energy integral (the first one in (13)) we have that
$$
K_1^2=h_1-\frac2{J_1+J_3}\chi_{(1)3}\Gamma_{(1)3}.
$$
Since $\tan l_1=\frac{M_{(1)1}}{M_{(1)2}}$, using first two
equations in (14), we have:
$$
\dot l_1=-2(J_{13}+J_{24})K_2\sin l_2+\frac{2\chi_{(1)3}c_1}{K_1^2}.
$$
Also from the second and third integral in (13), we have that
$$
K_1^2\Gamma_{(1)2}^2-2c_1M_{(1)2}\Gamma_{(1)2}+c_1^2-M_{(1)1}^2(1-\Gamma_{(1)3}^2)=0.
$$
Similarly, from equations (15), we get:
$$
\aligned
\dot\Gamma_{(2)3}^2&=4(J_1+J_3)^2\left[(1-\Gamma_{(2)3}^2)
(h_2-\frac2{J_1+J_3}\chi_{(2)3}\Gamma_{(2)3})-c_2^2\right]=
P_3(\Gamma_{(2)3}),\\
K_2^2&=h_2-\frac2{J_1+J_3}\chi_{(2)3}\Gamma_{(2)3},\\
\dot l_2&=-2(J_{13}-J_{24})K_1\sin l_1+\frac{2\chi_{(2)3}c_2}{K_2^2},\\
K_2^2\Gamma_{(2)2}^2&-2c_2M_{(2)2}\Gamma_{(2)2}+c_2^2-M_{(2)1}^2(1-\Gamma_{(2)3}^2)=0.
\endaligned
$$

From the previous considerations, we conclude that for complete
integration of the four-dimensional Hess-Appel'rot system one need
to solve a system of two differential equations (for $l_1$ and
$l_2$) of the first order and to calculate two elliptic integrals,
associated with elliptic curves $E_1$ and $E_2$ defined by
$$
E_i:\quad y^2=P_i(x)=8A_ix^3-4B_ix^2-8A_ix-4C_i,\ \ i=1,2
\tag {16}
$$
where
$$
A_i=(J_1+J_3)\chi_{(i)3}, \ B_i=(J_1+J_3)^2h_i,\
C_i=(J_1+J_3)^2(c_i^2-h_i).
$$
This is a typical situation for the Hess-Appel'rot systems that
additional integrations are required (see [36, 24, 15, 13]). Now
we pass to the algebro-geometric integration.

\

\

\centerline{\bf 6. Algebro-geometric integration}

\

Before analyzing spectral properties of the matrices $L(\lambda
)$, we will change the coordinates in order to diagonalize the
matrix $C$. In this new basis the matrices $L(\lambda)$ have the
form $\tilde L (\lambda )=U^{-1} L(\lambda ) U,$ where

$$
U=
\pmatrix
0 & 0 & \frac {i\sqrt 2}2 & \frac {\sqrt 2}2 \\
0 & 0 & \frac {\sqrt 2}2 & \frac {i\sqrt 2}2 \\
\frac {i\sqrt 2}2 & \frac {\sqrt 2}2 & 0 & 0 \\
\frac {\sqrt 2}2 & \frac {i\sqrt 2}2 & 0 & 0
\endpmatrix
$$
After straightforward calculations, we have

$$
\tilde L(\lambda )= \pmatrix
-i\Delta _{34} & 0 & -\beta _3^{*} -i\beta _4^{*} & i\beta _3-\beta _4 \\
0 &  i\Delta _{34} &  -i\beta _3^{*} -\beta _4^{*} & -\beta _3+i\beta _4\\
\beta _3-i\beta _4 & -i\beta _3 +\beta _4 & -i\Delta _{12} & 0 \\
i\beta _3^{*} +\beta _4^{*} & \beta _3^{*}+i\beta _4^{*} & 0 &
i\Delta _{12}
\endpmatrix
$$
where
$$
\aligned
\Delta _{12}&=\lambda ^2 C_{12}+\lambda M_{12}+\Gamma _{12},\\
\Delta _{34}&=\lambda ^2C_{34}+\lambda M_{34}+\Gamma _{34},
\endaligned
$$
$$
\xalignat 2
\beta _3&=x_3+\lambda
y_3, & x_3&=\frac 12 \left( \Gamma _{13}+i\Gamma _{23}\right),\\
 \beta _4 &=
x_4+\lambda y_4, & x_4&=\frac 12 \left( \Gamma _{14}+i\Gamma _{24}\right),\tag {17}\\
 \beta _3^{*}&=\bar x_3+\lambda \bar y_3, & y_3&=\frac 12 \left( M_{13}+
iM_{23}\right),\\
\beta _4^{*}&=\bar x_4+\lambda \bar y_4, & y_4&=\frac 12 \left( M_{14}+
iM_{24}\right).
\endxalignat
$$

Matrix $L(\lambda)$ is of the same form as the Lax matrix for the
Lagrange bitop [15, 16]. It is a quadratic polynomial in the
spectral parameter $\lambda $ with matrix coefficients. General
theories describing the isospectral deformations for polynomials
with matrix coefficients were developed by Dubrovin [17, 18] in
the middle of 70's and by Adler and van Moerbeke [1] a few years
later. Dubrovin's approach was based on the Baker-Akhiezer
function. Both approaches were   applied in rigid body problems
(see [32, 1] respectively).

But, as it was  shown in [16], none of these two theories can be
directly applied in  cases like this. Necessary  modifications
were suggested in [16], where a procedure of algebro-geometric
integration was presented. It is based on some nontrivial facts
from the theory of Prym varieties, such as the Mumford relation on
theta divisors of unramified double coverings and the
Mumford-Dalalyan theory (see [16, 35, 34, 14, 40, 5]).

Here, we are going to follow closely the procedure from [16], with
necessary changes, calculations and comments.

As usual,  we start with the spectral curve
$$
\Gamma : \, det\left( \tilde L(\lambda )-\mu \cdot 1 \right) =0.
$$
We have
$$
\Gamma : \, \mu ^4+\mu ^2\left( \Delta _{12}^2+\Delta _{34}^2+4\beta _3\beta _3
^{*}+4\beta _4\beta _4^{*}\right) +\left[ \Delta _{12}\Delta _{34}+2i(\beta
_3^{*}\beta _4-\beta _3\beta _4^{*})\right]^2=0.\tag {18}
$$
There is an involution
$$
\sigma:(\lambda,\mu)\rightarrow (\lambda, -\mu)
$$
of the curve $\Gamma$, which corresponds to the skew-symmetricity
of the matrix $L(\lambda)$. Denote the factor-curve by
$\Gamma_1=\Gamma/\sigma$.

\proclaim{Lemma 3} The curve $\Gamma_1$ is a smooth hyperelliptic
curve of the genus $g(\Gamma_1)=3$. The arithmetic genus of the
curve $\Gamma $ is $g_a(\Gamma) =9$.
\endproclaim
\demo{Proof} The curve
$$
\Gamma_1:\, u^2+P(\lambda)u+[Q(\lambda)]^2=0,
$$
is hyperelliptic, and its equation in the canonical form is:
$$
u_1^2=\frac{[P(\lambda)]^2}4- [Q(\lambda)]^2,
\tag{19}
$$
where $u_1=u+P(\lambda)/2$.
Since $\frac{[P(\lambda)]^2}4-[Q(\lambda)]^2$ is a polynomial of the
degree 8,
 the genus of the curve $\Gamma_1$ is $g(\Gamma_1)=3$.
$\Gamma$ is a double covering of $\Gamma_1$ and the ramification
divisor is of degree 8. According to the Riemann-Hurwitz formula,
 $g_a(\Gamma)=9$. \qed
\enddemo

\proclaim{Lemma 4} In generic case the spectral curve $\Gamma$ has four ordinary double
points $S_i, i=1,\dots , 4$. The genus of its normalization  $\tilde \Gamma$
is five.
\qed
\endproclaim
\demo {Proof} From the equations
$$
\frac{\partial p(\lambda, \mu)}{\partial \lambda}=0,\quad
\frac{\partial p(\lambda, \mu)}{\partial \mu}=0,
$$
where $p(\lambda, \mu)=det\left( \tilde L(\lambda )-\mu \cdot 1 \right)=
\mu ^4+\mu ^2P(\lambda)+[Q(\lambda)]^2$,
the double points are $S_k=(\lambda_k, 0), k=1,\dots,4$, where
$\lambda_k$ are zeroes of $Q(\lambda)$. Thus, $g(\tilde
\Gamma)=g_a(\Gamma)-4=5$. \qed
\enddemo

\proclaim{Lemma 5} Singular points $S_i$ of the curve $\Gamma$ are
fixed by $\sigma$. The involution $\sigma $ exchanges the two
branches of $\Gamma$ at $S_i$.
\endproclaim

\demo{Proof} Fixed points of the $\sigma$ are defined with
$\mu=0$, thus $S_i$ are fixed. Since their projections on
$\Gamma_1$ are smooth points, $\sigma$ exchanges the branches of
$\Gamma$, which are given by the equation:
$$
\mu^2=\frac{-P(\lambda)+\sqrt{P^2(\lambda)-4Q^2(\lambda)}}{2}.
$$
\qed
\enddemo

We start with  the well-known  eigen-problem

$$
\left(\frac {\partial }{\partial t}+\tilde A(\lambda )\right) \psi _k=0,\quad
\tilde L(\lambda )\psi _k=\mu _k \psi _k,
\tag{20}
$$
where $\psi _k$ are eigenvectors with eigenvalues $\mu _k$. Then
$\psi _k(t,\lambda )$ form a $4\times 4$ matrix with components
$\psi _k^i(t,\lambda )$. Denote by  $\varphi _i^k$ its inverse
matrix. Let us introduce
$$
g_j^i(t,(\lambda , \mu _k))=\psi _k^i(t,\lambda )\cdot \varphi _j^k(t,\lambda )
$$
(there is no summation on $k$)
or, in other words
$g(t)=\psi _k(t)\otimes \varphi(t) ^k.$

Matrix $g$ is of rank 1, and we have
$\partial \psi/\partial t=-\tilde{A}\psi,\quad \partial \varphi/
\partial t =\varphi \tilde{A},\quad \partial g/\partial t=
[g,\tilde{A}].$ We can consider vector-functions $\psi
_k(t,\lambda )= \left(\psi^1_k(t, \lambda),...,\psi^4_k(t,
\lambda)\right)^{T}$ as one function $\psi(t, (\lambda,
\mu))=\left(\psi^1(t,(\lambda, \mu)),..., \psi^4(t, (\lambda,
\mu))\right)^{T}$ on $\Gamma $ defined by $\psi ^i(t,(\lambda ,
\mu_k))$ $= \psi^i_k(t, \lambda)$. Similarly, we define
$\varphi(t,(\lambda,\mu))$. Relations for divisors of zeroes and
poles of functions $\psi ^i$ i $\varphi _i$ in the affine part of
$\Gamma $ are:

$$
\left( g^i_j\right)_a=d_j(t)+d^i(t)-D_r-D'_s,
\tag {21}
$$
where $d_j(t)$ is divisor of zeroes of $\psi_j$, divisor $d^i(t)$ is
divisor of zeroes of $\varphi^i$, $D_r$ is the ramification divisor over $\lambda$ plane (see
[17]), $D'_s$ is some subdivisor of $D_s$ divisor of singular points defined by (21).
One can easily calculate $deg \,D_r=16, deg D_s=8$.

Matrix elements $g_j^i(t,(\lambda ,\mu _k))$  are meromorphic
functions on $\Gamma $. We need their asymptotics  in
neighbourhoods of points $P_k$, which cover the point $\lambda
=\infty $. Let $\tilde \psi _k$ be the eigenvector of the matrix
$\tilde L(\lambda )$ normalized in $P_k$ by the condition $\tilde
\psi _k^k=1$, and let $\tilde \varphi_i^k$ be the inverse matrix
for $\tilde \psi _k^i$. We will also use another decomposition of
matrix elements of $g$: $g_j^i=\psi _k^i\varphi _j^k=\tilde \psi
_k^i\tilde \varphi ^k_j.$ It is an immediate consequence of
proportionality of the vectors $ \psi _k$ and $\tilde \psi _k$ ($
\varphi^k$ and $\tilde \varphi^k$).

\proclaim {Lemma 6} \item{{\rm a)}} Matrix $g$ has the following
representation
$$
  g=\frac {\mu ^3 + a_1\mu ^2 + a_2\mu + a_3}{\partial p(\lambda, \mu)/\partial \mu},
$$
where $a_1=L, a_2=P\cdot 1+L^2, a_3=PL+L^3.$

\item{{\rm b)}} For the Lax matrix $L$ and $\lambda _i$ such that
$Q(\lambda_i)=0,$ it holds $a_3=0.$
\endproclaim

The proof of the Lemma follows from [17] and straightforward
calculation. From the part (a) one can see that $g$ could have
poles in singular points of the spectral curve. But, from (b) we
have

\proclaim {Corollary 1} The matrix $g$  has no poles in singular
points of the curve $\Gamma $.
\endproclaim

So, from now on, taking Corollary 1  into account, we will
consider all functions in this section as functions on the
normalization $\tilde \Gamma $ of the curve $\Gamma $.

Since the functions $\tilde \psi _k^i$ and $\tilde \varphi _j^k$
are meromorphic in neighbourhoods of points $P_k$, their
asymptotics can be calculated by expanding  $\tilde \psi _k$ as a
power series in $\lambda ^{-1}$ in a neighbourhood of the point
$\lambda =\infty $ around the vector $e_k$, where $e_k^i=\delta
_k^i$. We get
$$
\aligned
&\left( \tilde C+\frac {\tilde M}{\lambda }+\frac {\tilde \Gamma }
{\lambda ^2}\right)\left( e_i+\frac {u_i}{\lambda }+\frac {v_i}{\lambda ^2}+
\frac {w_i}{\lambda ^3}+\dots \right)\\
&=\left( \tilde C_{ii}+\frac {b_i}{\lambda }+\frac {d_i}{\lambda ^2}+
\frac {h_i}{\lambda ^3}+\dots \right) \left( e_i+\frac {u_i}{\lambda }+
\frac {v_i}{\lambda ^2}+\frac {w_i}{\lambda ^3}+\dots \right),
\endaligned
\tag {22}
$$
where matrices $\tilde C, \tilde M$ and  $\tilde \Gamma $ are
defined by $\tilde L(\lambda )=\lambda ^2\tilde C+\lambda \tilde
M+\tilde \Gamma.$ Comparing the same powers of $\lambda$, from
(22) we get
$$
\aligned
(u_i)_i&=0,\quad (v_i)_i=0,\quad (w_i)_i=0\\
(u_i)_j&=\frac {\tilde M_{ji}}{\tilde C_{ii}-\tilde C_{jj}} \qquad j\ne i\\
(v_i)_j&=\frac 1{\tilde C_{ii}-\tilde C_{jj}}\left( \sum_{k\ne i}\frac
{\tilde M_{jk}\tilde M_{ki}}{\tilde C_{ii}-\tilde C_{kk}}-\frac {\tilde M_{ii}
\tilde M_{ji}}{\tilde C_{ii}-\tilde C_{jj}}+\tilde \Gamma _{ji}\right) \\
(w_i)_j&=\frac 1{C_i-C_j}\left[ \sum_{k\ne i}\tilde M_{jk}(v_i)_k+\sum_{k\ne i}
\tilde \Gamma _{jk}(u_i)_k-b_i(v_i)_j-d_i(u_i)_j\right]\\
b_i&=\tilde M_{ii},\quad
d_i=\sum_{k\ne i}\frac {\tilde M_{ik}\tilde M_{ki}}{\tilde C_{ii}-
\tilde C_{kk}}+\tilde\Gamma _{ii},\quad
h_i=\sum_{k\ne i}\tilde M_{ik}(v_i)_k+\sum_{k\ne i}
\tilde \Gamma _{jk}(u_i)_k
\endaligned \tag 23
$$
So, the matrix  $\tilde \psi =\{ \tilde \psi _k^i\}$ in a
neighbourhood of $\lambda =\infty $ has the form:
$$
\tilde \psi =1+\frac u{\lambda }+\frac v{\lambda ^2}+\frac w{\lambda ^3} +
O\left( \frac 1{\lambda ^3}\right).
\tag{24}
$$

Denote by $\tilde d_j$ and $\tilde d^i$ the following divisors:
$$
\aligned
&\tilde d_1=d_1+P_2,\quad \tilde d_2=d_2+P_1,\quad \tilde d_3=d_3+P_4,\quad
\tilde d_4 =d_4+P_3,\\
&\tilde d^1=d^1+P_2,\quad \tilde d^2=d^2+P_1,\quad \tilde d^3=d^3+P_4,\quad
\tilde d^4 =d^4+P_3.
\endaligned
$$
Analyzing the behavior of matrix $g$ around points $P_k$, as in
[16], we get

\proclaim{Proposition 3} \item{{\rm a)}} Divisors of matrix
elements of $g$ are
$$
\left( g_j^i\right)=\tilde d^i+\tilde d_j-D_r+2\left(
P_1+P_2+P_3+P_4\right)-P_i-P_j
\tag{25}
$$
\item{{\rm b)}} Divisors $ \tilde d_i, \tilde d^j$ are of the same
degree
$$
deg\, \tilde d_i=\deg \, \tilde d^j=5.
$$
\endproclaim

Let us denote by $\Phi(t,\lambda)$ the normalized fundamental
solution of
$$
\left(\frac{\partial}{\partial
t}+\tilde{A}(\lambda)\right)\Phi(t,\lambda)=0,\quad \Phi(\tau)=1.
$$
Then, if we introduce the Baker-Akhiezer functions
$$
\hat{\psi}^i(t,\tau,(\lambda,\mu_k))=\sum_s\Phi^i_s(t,\lambda)h^s(\tau,(\lambda,\mu_k))
\tag{26}
$$
where $h^s$ are eigen-vectors of $L(\lambda)$ normalized by the
condition $\sum_s h^s(t,(\lambda,\mu_k))=1$, it follows that:
$$
\hat{\psi}^i(t,\tau,(\lambda,\mu_k))=\sum_s\Phi^i_s(t,\lambda)
\frac{\psi^s_k(\tau,\lambda)}{\sum_l\psi^l_k(\tau,\lambda)}=
\frac{\psi^i_k(t,\lambda)}{\sum_l\psi^l_k(\tau,\lambda)}.
\tag{27}
$$
\proclaim{Proposition 4} Functions $\hat{\psi}^i$ satisfy the
following properties \item {{\rm a)}} In the affine part of
$\tilde \Gamma$, the function $\hat{\psi}^i$ has 4 time dependent
zeroes which belong to the divisor $d^i(t)$ defined by formula
{\rm(21)}, and 8 time independent poles, i.e.
$$
\left(\hat{\psi}^i(t,\tau,(\lambda,\mu_k))\right)_a=d^i(t)-\bar{\Cal D},
\qquad \deg\bar{\Cal D}=8.
$$
\item {{\rm b)}} In points $P_k$, functions $\hat{\psi}^i$ have
essential singularities as follows:
$$
\hat{\psi}^i(t, \tau , (\lambda , \mu )) = exp \, \left[ -(t-\tau
) R_k-iF_k)\right] \hat{\alpha}^i(t,\tau,(\lambda,\mu))
$$
where $R_k$ and $F_k$ are:
$$
\aligned &R_1=i\left(\frac{\chi_{34}}{z}\right), \;R_2=-R_1,\;
R_3=i\left(\frac{\chi_{12}}{z}\right),\; R_4=-R_3,\\
F_1&=\left(\int _{\tau}^t\Omega_{34}dt\right),\; F_2=-F_1,\;
F_3=\left(\int _{\tau}^t\Omega_{12}dt\right), F_4=-F_3
\endaligned
$$
and $\hat{\alpha}^i$ are holomorphic in a neighbourhood of $P_k$,
$$
\hat{\alpha}^i(\tau,\tau,(\lambda,\mu))=h^i(\tau,(\lambda,\mu)),\quad
\hat{\alpha}^i(t,\tau,P_k)=\delta_i^k+\tilde v^i_k(t)z+O(z^2),
$$
with
$$
\tilde v^i_k=\frac{\tilde{M_{ki}}}{\tilde{C_{ii}}-\tilde{C_{kk}}}.
\tag {28}
$$
\endproclaim

Proof repeats the demonstration of Proposition 5 in [16].

Let us denote by $\hat{\psi}^i_{LB}$ the Baker-Akhiezer function
for the Lagrange bitop from [16] with analytical properties as in
Proposition 4 a) above and with asymptotics given by:

at points $P_k$, functions $\hat{\psi}^i_{LB}$ have essential
singularities as follows:
$$
\hat{\psi}^i_{LB}(t, \tau , (\lambda , \mu )) = exp \, \left[
-(t-\tau ) R_k\right] \hat{\alpha}^i_{LB}(t,\tau,(\lambda,\mu)),
$$
where $R_k$ are given with
$$
R_1=i\left(\frac{\chi_{34}}{z}\right),
\;R_2=-R_1,\;
R_3=i\left(\frac{\chi_{12}}{z}\right),\;
R_4=-R_3,
$$
and $\hat{\alpha}^i_{LB}$ are holomorphic in a neighbourhood of
$P_k$,
$$
\hat{\alpha}^i_{LB}(\tau,\tau,(\lambda,\mu))=h^i_{LB}(\tau,(\lambda,\mu)),\quad
\hat{\alpha}^i_{LB}(t,\tau,P_k)=\delta_i^k+\tilde v^i_k(t)z+O(z^2).
$$
From the Proposition 4 and from Proposition 5 of [16],
 we have

\proclaim {Corollary 2} A relationship between the data of
generalized Hess-Appel'rot problem and the Lagrange bitop are
given by: \item {{\rm a)}}
$$
\hat {\psi}^k_{HA}:=\hat {\psi}^k = \exp (i F_k)\hat {\psi}^k_{LB},
\quad k=1,\dots ,4;
$$
\item {{\rm b)}}
$$
v^k_{j HA}:=\tilde v^k_j = \exp (i (F_k+F_j))v^k_j, \quad
k,j=1,\dots, 4. \tag {29}
$$
\endproclaim
($v^k_j$ we will also denote as $v^k_{j LB}$.)

\

\

\centerline {\bf 7. A Prym variety}

\

Let us recall that $d^j(t)$ is divisor defined in (21).
\proclaim{Lemma 7} On the Jacobian $\Jac(\tilde \Gamma)$ the
following relation takes place :
$$
\Cal A(d^j(t)+\sigma d^j(t))=\Cal A(d^j(\tau)+\sigma d^j(\tau))
$$
where $\Cal A$ is the Abel map from the curve $\tilde \Gamma$ to
$\Jac(\tilde \Gamma)$, and $\sigma$ is involution on $\tilde \Gamma$.
\endproclaim

The proof is the same as the one of the corresponding Lemma in
[16].

From the previous Lemma, we see that vectors $\Cal A (d^i(t))$
belong to some translation of a Prym variety $\Pi = Prym(\tilde
\Gamma|\Gamma_1)$. More details concerning Prym varieties one can
find in [41, 40, 21, 9, 34, 35, 5, 8]. A natural question arises
to compare two-dimensional tori $\Pi$ and $E_1\times E_2$, where
elliptic curves $E_i$ are defined by (16).

\

Together with the curve $\Gamma_1, $ one can consider
curves $\Cal C_1$ and $\Cal C_2$ defined by the equations
$$
\Cal C_1: v^2=\frac{P(\lambda)}2 + Q(\lambda),\quad
\Cal C_2: v^2=\frac{P(\lambda)}2 - Q(\lambda).
\tag{30}
$$

\proclaim{Lemma 8} Curves $E_i$ defined by {\rm (16)} are
Jacobians of curves $\Cal C_i$ given by {\rm (30)}.
\endproclaim
\demo{Proof} Follows  by a straightforward calculation.
\enddemo \qed

Since the curve $ \Gamma_1$ is hyper-elliptic, in a study of the
Prym variety $\Pi$ the Mumford-Dalalyan theory can be applied (see
[14, 34, 40]). Thus, the previous Lemma allows us to use the
following Theorem from [16].

\proclaim {Theorem 3} \item{{\rm a)}} The Prymian $\Pi$
is isomorphic to the product of curves $E_i$:
$$
\Pi = Jac (\Cal C_1)\times Jac (\Cal C_2).
$$
\item {{\rm b)}} The curve $\tilde \Gamma $ is the
desingularization of $\Gamma_1\times _{\Bbb P^1} \Cal C_2$ and
$\Cal C_1 \times _{\Bbb P^1} \Gamma_1.$ \item {{\rm c)}} The
canonical polarization divisor $\Xi$ of $\Pi$ satisfies
$$
\Xi=E_1\times \Theta_2 + \Theta _1\times E_2,
$$
where $\Theta _i $ is the theta-divisor of $E_i$.

\endproclaim

Theorem 3 explains the connection between the curves $E_1, E_2$
and the Prym variety $\Pi$. Further analysis of properties of Prym
varieties necessary for understanding the dynamics of the Lagrange
bitop will be done in the next section.

\

\centerline {\bf 8.  Isoholomorphisity condition, Mumford's
relation and} \centerline {\bf solutions for $v^k_{j LB}$}

\

We saw that integration of the four-dimensional Hess-Appel'rot
system is partially reduced to solutions of the Lagrange bitop.
Now, we are goint to give the explicit formulae for the
Baker-Akhiezer function for the Lagrange bitop, obtained in [16].
According to Proposition 4, the Baker-Akhiezer function $\Psi$
satisfies usual conditions of normalized (n=)4-point function on a
curve of genus $g=5$ with the divisor $\bar \Cal D$ of degree
$\deg \bar \Cal D= g+n-1=8$, see [19, 18]. By the general theory,
it should determine the whole dynamics uniquely.

Let us consider the differentials $\Omega^i_j=g_{ij}d\lambda,\quad
i,j=1,\dots , 4.$ In the case of general position it was proved by
Dubrovin  that $\Omega^i_j$ is a meromorphic differential having
poles at $P_i$ and $P_j$, with residues $v^i_j$ and $-v^j_i$
respectively. But here we have

\proclaim{Proposition 5} {\rm[16]} Differentials
 $\Omega^1_2,\; \Omega^2_1,\; \Omega^3_4,\; \Omega^4_3$
 are holomorphic during the whole evolution.
 \endproclaim
The proof is based on the fact that from the conditions
$L^1_2=L^2_1=L^3_4=L^4_3=0$ it follows that
$$
v^1_2=v^2_1=v^3_4=v^4_3=0.
\tag {31}
$$
(For more details see [16]). We can say that the condition
$L^1_2=L^2_1=L^3_4=L^4_3=0$ implies {\it isoholomorphicity}. Let
us recall the general formulae for $v$ from [18]:
$$
v^i_j=\frac{\lambda_i \theta (A(P_i)-A(P_j)+tU+z_0)} {\lambda_j
\theta (tU+z_0) \epsilon (P_i,P_j)},\quad  i\ne j,
\tag {32}
$$
where $U =\sum x^{(k)}U^{(k)}$ is a certain linear combination of
$b$-periods $U^{(i)}$ of differentials of the second kind
$\Omega^{(1)}_{P_i}$, which have a pole of order two at $P_i$;
$\lambda_i$ are nonzero scalars, and
$$
\epsilon (P_i,P_j):=\frac {\theta [\nu ](A(P_i-P_j))}{(-\partial _{U^{(i)}}\theta [\nu ](0))^{1/2} (-\partial _{U^{(j)}}\theta [\nu ](0))^{1/2})}.
$$
(Here $\nu $ is an arbitrary odd non-degenerate characteristic.)
Thus, from (32) we get:

{\it Holomorphicity of some of the differentials $\Omega^i_j$
implies that the theta divisor of the spectral curve contains some
tori.}

\

In a case when the spectral curve is a double unramified covering
$$
\pi: \tilde \Gamma \rightarrow \Gamma _1;
$$
with $g(\Gamma _1)=g, \quad g(\tilde \Gamma )=2g-1$, as we have
here (assuming that $\tilde\Gamma$ is the normalization of the spectral curve $\Gamma$),
it is really satisfied that the theta divisor contains a
torus, see [35]. Let us denote by $\Pi^-$ the set
$$
   \Pi^-=\left\{ L\in Pic^{2g-2} \tilde \Gamma | Nm L = K_{\Gamma 1}, h^0(L) ~
   \text {is odd} \right\},
$$
where $K_{\Gamma _1}$ is the canonical class of the curve $\Gamma
_1$ and $Nm: Pic \tilde \Gamma \rightarrow Pic \Gamma _1$ is the
norm map, see [35, 40] for details. For us,  it is crucial that
$\Pi^-$ is a translate of the Prym variety $\Pi$ and that
Mumford's relation ([35]) holds:
$$
\Pi^-\subset \Theta _{\tilde \Gamma}.
\tag {33}
$$
Let us denote
$$
U= i(\chi_{34} U^{(1)}-\chi_{34} U^{(2)}+ \chi_{12} U^{(3)}-\chi_{12} U^{(4)}),
\tag {34}
$$
where $U^{(i)}$ is the vector of $\tilde b$-periods of the
differential of the second kind $\Omega^{(1)}_{P_i}$, which is
normalized by the condition that $\tilde a$-periods are zero. We
suppose here that the cycles $\tilde a, \tilde b$ on the curve
$\tilde \Gamma$ and $a, b$ on $\Gamma_1$ are chosen  to correspond
to the involution $\sigma $ and the projection $\pi$, see [5, 40]:
$$
\pi (\tilde a_0)=a_0; \quad \pi(\tilde b_0)= 2 b_0,\quad
\sigma (\tilde a_k)= \tilde a_{k+2}, \quad k=1,2.
$$
The basis of normalized holomorphic differentials $[u_0,\dots ,u_5]$ on $\tilde \Gamma $ and $[v_0, v_1,v_2]$ on $\Gamma_1$  are chosen such that
$$
\pi^*(v_0)=u_0,\quad \pi^*(v_i)=u_i+\sigma(u_i)=u_i+u_{i+2},\quad
i=1,2.
$$
Now we have

\proclaim{Theorem 4} {\rm[16]} \item{{\rm a)}} If vector $z_0$ in
{\rm(32)} corresponds to the translation of the Prym variety $\Pi$
to $\Pi^-$, and vector $U$ is defined by {\rm(34)} then conditions
{\rm(31)} are satisfied.

\item{{\rm b)}} The explicit formula for $z_0$ is
$$
z_0=\frac{1}{2}(\hat \tau_{00},\hat \tau_{01},\hat \tau_{02},\hat \tau_{01},\hat \tau_{02}),\quad
\hat \tau_{0i}=\int_{\tilde b_0} u_i, \quad i=0, 1, 2.
\tag {35}
$$
\endproclaim

Formulae for scalars $\lambda_i$ from (32) will be given later in
this section.

\

The evolution on the Jacobian of the spectral curve $\Jac (\tilde
\Gamma)$ gives a possibility to reconstruct the evolution of Lax
matrix $L(\lambda)$ only up to a conjugation by diagonal matrices.
To overcome this problem, we are going to consider, together with
Dubrovin, a generalized Jacobian, obtained by gluing together the
infinite points. Those points are $P_1, P_2, P_3, P_4$ and the
corresponding Jacobian will be denoted by $\Jac(\tilde \Gamma|
\left \{ P_1, P_2, P_3, P_4\right \}).$

The generalized Jacobian can be understood as a set of classes of
relative equivalence among the divisors on $\tilde \Gamma $ of a
certain degree. Two divisors of the same degree  $D_1$ and $D_2$
are called {\it equivalent relative to points} $P_1, P_2, P_3, P_4
$ if there exists a meromorphic function $f$ on $\tilde \Gamma $
such that $(f)=D_1-D_2$ and $f(P_1)=f(P_2)=f(P_3)=f(P_4)$.

The generalized Abel map is defined with
$$
\tilde A(P)=(A(P),\lambda_1 (P),...,\lambda_4(P)),\quad
\lambda_i(P)=\exp\int_{P_0}^P\Omega_{P_iQ_0}, i=1,...,4,
$$
where $A$ is the standard Abel map. Here $\Omega_{P_iQ_0}$ denotes
the normalized differential of the third kind, with poles at $P_i$
and at an arbitrary fixed point $Q_0$.

Then the generalized Abel theorem (see [21]) can be formulated as

\proclaim {Lemma 9 (the generalized Abel theorem)} Divisors $D_1$
and $D_2$ are equivalent relative $P_1, P_2, P_3, P_4$ if and only
if there exist integer-valued vectors $N, M$ such that
$$
\aligned
A(D_1)&= A(D_2) + 2\pi N +BM,\\
\lambda_j (D_1)& = c \lambda_j (D_2)\exp(M,A(D_2)),\,\, j=1,...,4
\endaligned
$$
where $c$ is some constant and $B$ is the period matrix of the curve $\tilde \Gamma$.
\endproclaim

A generalized Jacobi inverse problem can be formulated as a
question of finding, for given $z$, points $Q_1,\dots, Q_8$ such
that
$$
\aligned
\sum_1^8A(Q_i) - \sum_2^4A(P_i)&=z+K,\\
\lambda_j=c \exp\sum_{s=1}^8\int_{P_0}^{Q_s}\Omega _{P_jQ_0} &+
\kappa_j, \,\,j=1,...4,
\endaligned
$$
where $K$ is the Riemann constant and constants $\kappa_j$ depend
on $\tilde \Gamma$, points $P_1, P_2, P_3, P_4$ and the choice of
local parameters around them.

We will denote by $Q_s$ the points which belong to the divisor
$\bar \Cal D$ from Proposition 4, and by $E$ the prime form from
[21]. Then we have

\proclaim{Proposition 6} Scalars $\lambda_j$  from formula
{\rm(32)} are given with
$$
\lambda_j=\lambda_j^0 \exp \sum_{k\ne j}ix^{(k)}\gamma_j^k,\qquad
 \lambda_j^0=c \exp\sum_{s=1}^8\int_{P_0}^{Q_s}\Omega _{P_jQ_0} + \kappa_j,
$$
where $\vec x= (x^{(1)},\dots,
x^{(4)})=t(\chi_{34},-\chi_{34},\chi_{12},-\chi_{12})$ and
$$
\gamma_i^j=\frac{d}{dk_j^{-1}}ln E(P_i,P)|_{P=P_j}.
$$
($k_j^{-1}$ is a local parameter around $P_j$.)
\endproclaim

\

To give formulae for the Baker-Akhiezer function, we need some
notations. Let
$$
\alpha ^j(\vec x)=\exp[i\sum \tilde \gamma _m^jx^{(m)}]\frac
{\theta (z_0)}{\theta (i\sum x^{(k)}U^{(k)} +z_0)},
$$
where
$$
\tilde \gamma ^j_m=\int _{P_0}^{P_j}\Omega^{(1)}_{P_m}, \quad m\ne j,
$$
and $\tilde \gamma^m_m$ is defined by the expansion
$$
\int _{P_0}^{P}\Omega^{(1)}_{P_m}=-k_m+\tilde \gamma^m_m + O(k_m^{-1}),\quad P \rightarrow P_m.
$$
Denote
$$
\phi ^j(\vec x, P)= \alpha ^j(\vec x) \exp(-i\int _{P_0}^P\sum
x^{(m)}\Omega^{(1)}_{P_m})\frac {\theta (A(P)-A(P_j)-i\sum
x^{(k)}U^{(k)}-z_0)}{\theta (A(P)-A(P_j)-z_0)}.
$$
Finally we come to
\proclaim{Proposition 7} {\rm[16]} The
Baker-Akhiezer function is given by
$$
\psi ^j(\vec x, P)=\phi ^j(\vec x, P)\frac {\lambda_j^0\frac {\theta (A(P-P_j)-z_0)}{\epsilon (P, P_j)}}{\sum_{k=1}^4 \lambda _k^0 \frac {\theta (A(P-P_k)-z_0)}{\epsilon (P, P_k)}}, \quad j=1,\dots ,4,
$$
where $z_0$ is given by {\rm(35)}.
\endproclaim

\

\centerline {\bf 9. The restrictively integrable part --}
\centerline{\bf equations for the functions $F_i, i=1,\dots,4$}

\

Let us denote
$$
\aligned
\phi_1&:=F_1+F_3,\quad  \phi_2:=F_1-F_3;\\
\endaligned
$$
and also
$$
\aligned
N_1&:=M_{14}-M_{23}, \quad N_3:=M_{24}-M_{13}, \\
N_2&:=-M_{24}-M_{13}, \quad N_4:=M_{14}+M_{23}.\\
\endaligned
$$
From (29) we have
$$
\varphi_1 :=arg(v^1_{3HA})=\phi_1+\alpha_1(t),\quad \varphi_2
:=arg(v^1_{4HA})=\phi_2+\alpha_2(t),
$$
where $\alpha_1(t)=arg(v^1_{3LB})$ and
$\alpha_2(t)=arg(v^1_{4LB})$ are known function of time. Let us
denote $u_i=\tan\varphi_i$.

 Basic relationships among those quantities are
given in the next proposition.

\proclaim {Proposition 8} The following relations take place \item
{{\rm a)}}
$$
u_1=\frac {N_1}{N_2},\quad u_2=-\frac {N_3}{N_4}; \tag {36}
$$
\item {{\rm b)}}
$$
N_1=-2|v^1_{3LB}(\tilde {C_{11}}-\tilde {C_{33}})|\sin \varphi_1,
\quad N_4=2|v^1_{4LB}(\tilde {C_{11}}-\tilde {C_{44}}) |\cos
\varphi_2; \tag {37}
$$
\item {{\rm c)}}
$$
\dot \phi_1=N_1(J_{24}+J_{13}),\quad \dot
\phi_2=-N_4(J_{24}-J_{13}). \tag {38}
$$
\endproclaim

\demo {Proof} a) follows from the formulae for $\tilde L$ (28) and
(29). Part b) also uses  Corollary 2 b, Proposition 4 b. Note that
from the condition $\Omega =JM+MJ $ and the invariant relations we
have:
$$
\aligned
\Omega_{12}&=M_{14}J_{24}+M_{32}J_{13};\\
\Omega_{34}&=M_{32}J_{24}+M_{14}J_{13}.\\
\endaligned
$$
From the last relation and the definition of functions $F_i$ from
the Proposition 4, c) follows. \qed
\enddemo

Using formulae (37), (38) we get
$$
\aligned
\dot\phi_1&=-2(J_{24}+J_{13})|(\tilde C_{11}-\tilde
C_{33})v^1_{3LB}|\sin(\phi_1+\alpha_1(t))\\
\dot\phi_2&=-2(J_{24}-J_{13})|(\tilde C_{11}-\tilde
C_{44})v^1_{4LB}|\cos(\phi_2+\alpha_2(t))
\endaligned
$$

\

\

\centerline{\bf 10. Restrictive integrability in an abstract
Poisson algebra settings.}
\centerline{\bf Bihamiltonian
structures for the Lagrange bitop}
\centerline{\bf and
$n$-dimensional Lagrange top}

\

From the analysis given in this paper, it follows that the
Hess-Appel'rot system and its generalizations can be understood as
natural examples of the following, more abstract situation.

Suppose a Poisson manifold $(M^{2n}, \{\cdot,\cdot\})$ is given,
together with $k+1$ functions $H, f_1,\dots, f_k \in
C^{\infty}(M)$, such that

\item {\bf (A1)}
$$
\{H, f_i\}= \sum_{j=1}^k a_{ij}f_j, \quad a_{ij}\in C^{\infty}(M),\quad i, j=1,\dots ,k;
$$
\item {\bf (A2)}
$$
\{f_i,f_j\}=0,\quad i,j=1,\dots ,k.
$$
The Hamiltonian system $(M^n, H)$ will be called  {\it
restrictively integrable}, if it satisfies the axioms (A1-A2).

A more general case can be obtained by replacing condition (A2)
with \item{\bf (A2')}
$$
\{f_i,f_j\}=\sum_{l=1}^k d_{ij}^l f_l,\quad d_{ij}^l=const,\ \
i,j=1,...,k.
$$
In this case, the algebra of invariant relations is a
noncommutative Lie algebra.

Starting from the Hamiltonian system $(M, H_0)$ with $k$ integrals
in involution $f_1,\dots, f_k$, choosing functions $b_j\in
C^{\infty}(M), \quad j=1,\dots ,k$, one comes to a restrictively
integrable system: \item {(HP)} {\bf Hamiltonian perturbation}

The system $(M, H)$ where
$$
H=H_0 + \sum _{j=1}^kb_j f_j,
$$
will be called a Hamiltonian perturbation. It satisfies (A1) with
$$
a_{ij}=\{b_j, f_i\}, \quad i,j=1,\dots, k.
$$
Natural question is the converse one: when a restrictively
integrable system is of the form (HP)?

Denote
$$
c_{il}^j:=\{a_{ij}, f_l\}, \quad i, j, l= 1\dots, k.
$$
From the Jacobi identity, and involutivity of functions $f_i$ we
get compatibility conditions.

\proclaim{Proposition 9} If a restrictively integrable system
which satisfies the axioms {\rm (A1-A2)} is of the form {\rm
(HP)}, then
$$
c_{il}^j=c_{li}^j, \quad i, j, l= 1\dots, k.
$$
\endproclaim

If in Proposition 9 we replace axiom A2 with A2' then $c_{il}^j$
should satisfy
$$
c_{il}^j=c_{li}^j+\sum_md_{il}^ma_{jm}.
$$

\

\

A three-dimensional Lagrange top is defined by the Hamiltonian:
$$
H_L=\frac 12
\left(\frac{M_1^2+M_2^2}{I_1}+\frac{M_3^2}{I_3}\right)+z_0\Gamma_3,
$$
according to the standard Poisson structure
$$
\{M_i, M_j\}_1=-\epsilon_{ijk}M_k,\,\,
\{M_i,\Gamma_j\}_1=-\epsilon_{ijk}\Gamma_k,\,\,\{\Gamma_i,
\Gamma_j\}=0
$$
on the Lie algebra $e(3)$. It is also well-known that
three-dimensional Lagrange top is Hamiltonian in another Poisson
structure, compatible with first one. This structure is defined
by:
$$
\{\Gamma_i,\Gamma_j\}_2=-\epsilon_{ijk}\Gamma_k,\,\,\{M_1, M_2\}_2=1,
$$
and the corresponding Hamiltonian is:
$$
\tilde H_L=(a-1)M_3\left(\frac12
(M_1^2+M_2^2)+\Gamma_3\right)+M_1\Gamma_1+M_2\Gamma_2+M_3\Gamma_3
$$
where $I_1=1, I_3=a, z_0=1$.

Casimir functions in the second structure are
$\Gamma_1^2+\Gamma_2^2+\Gamma_3^2$ and $M_3$.

 Let us observe that the
Hamiltonian for the three-dimensional Hess-Appel'rot case is a
quadratic deformation of Hamiltonian $H_L$ of the Lagrange top:
$$
H_{HA}=H_L+J_{13}M_1M_3.
$$
The function $M_3$, which gives the invariant relation for the
Hess-Appel'rot case, is a Casimir function of the second Poisson
structure.

Having this observation in mind, next we are going to prove that
the Lagrange bitop and the $n$-dimensional Lagrange top are also
bihamiltonian systems.

The standard Poisson structure on the semi-direct product $so(4)\times so(4)$
is:
$$
\{M_{ij}, M_{jk}\}_1=-M_{ik},\quad \{M_{ij},
\Gamma_{jk}\}_1=-\Gamma_{ik},\quad \{\Gamma_{ij},\Gamma_{kl}\}_1=0.
$$
Now let us introduce a new Poisson structure as follows:
$$
\aligned
\{\Gamma_{ij},\Gamma_{jk}\}_2&=-\Gamma_{ik},\quad
\{M_{ij},\Gamma_{kl}\}_2=0,\\
\{M_{13}, M_{23}\}_2&=-\chi_{12},\quad \{M_{14},
M_{24}\}_2=-\chi_{12},\\
\{M_{13}, M_{14}\}_2&=-\chi_{34},\quad \{M_{23}, M_{24}\}_2=-\chi_{34}
\endaligned
\tag{39}
$$

Casimir functions in this structure are $M_{12}$, $M_{34}$,
$\Gamma_{12}^2+\Gamma_{13}^2+\Gamma_{14}^2+\Gamma_{23}^2+\Gamma_{24}^2+\Gamma_{34}^2$,
and
$\Gamma_{12}\Gamma_{34}+\Gamma_{23}\Gamma_{14}-\Gamma_{13}\Gamma_{24}$.

 \proclaim{Proposition 10} The Poisson structure {\rm (39)} is
compatible with the standard one.
\endproclaim

\demo{Proof} Two Poisson structures, defined with antisymmetric
matrices $A$ and $B$, are compatible if their Shouten bracket,
defined by:
$$
[A, B]_{ijk}=\sum_s \left(\frac{\partial A^{ij}}{\partial
x^s}B^{sk}+\frac{\partial B^{ij}}{\partial
x^s}A^{sk}\right)+\text{cyclic for }i,j,k,
$$
vanishes (see [24]). Proof follows by direct calculation. \qed
\enddemo

%Four-dimensional Lagrange system, in metric $M=I\Omega+\Omega I$,
%where $I$ is symmetric matrix, introduced by Ratiu in [32] is
%defined with Hamiltonian
%$$
%H_L=\frac12\left(\frac{M_{12}^2}{2I_1}+\frac{M_{13}^2}{I_1+I_3}+\frac{M_{14}^2}{I_1+I_3}+
%\frac{M_{23}^2}{I_1+I_3}+\frac{M_{24}^2}{I_1+I_3}+\frac{M_{34}^2}{2I_3}\right)
%+\chi_{12}\Gamma_{12}
%$$
%and it is completely integrable.

%We can assume that $I_1=a, I_3=1-a, \chi_{12}=1$.

%If metric is given with $\Omega=JM+MJ$, then we can also define
%Lagrange top with Hamiltonian
%$$
%\aligned
%H_{L1}&=\frac12(2J_1M_{12}^2+(J_1+J_3)M_{13}^2+(J_1+J_3)M_{14}^2+\\
%&(J_1+J_3)M_{23}^2+(J_1+J_3)M_{24}^2+2J_3M_{34}^2)
%+\chi_{12}\Gamma_{12}
%\endaligned
%$$

In the metric $\Omega=JM+MJ$,
where $J=\diag(J_1, J_1, J_3, J_3)$, Hamiltonian function of the
Lagrange bitop in the standard Poisson structure is:
$$
\aligned
H_{LB}&=\frac12(2J_1M_{12}^2+(J_1+J_3)M_{13}^2+(J_1+J_3)M_{14}^2+\\
&(J_1+J_3)M_{23}^2+(J_1+J_3)M_{24}^2+2J_3M_{34}^2)
+\chi_{12}\Gamma_{12}+\chi_{34}\Gamma_{34}.
\endaligned
$$

Let us assume that $J_1=a$, $J_3=1-a$.

\proclaim{Proposition 11} The Lagrange bitop defined in the first
Poisson structure by the Hamiltonian $H_{LB}$ is a Hamiltonian
system in the second Poisson structure (39) with the Hamiltonian:
$$
\aligned  \tilde
H_{LB}&=\\
&\frac{(2a-1)(\chi_{12}M_{12}+\chi_{34}M_{34})}{\chi_{12}^2-\chi_{34}^2}
\left(\frac{M_{13}^2+ M_{14}^2+
M_{23}^2+M_{24}^2}2+\chi_{12}\Gamma_{12}+\chi_{34}\Gamma_{34}\right)\\
&+\frac{(1-2a)(\chi_{12}M_{34}+\chi_{34}M_{12})}{\chi_{12}^2-\chi_{34}^2}
(M_{23}M_{14}-M_{13}M_{24}+\chi_{12}\Gamma_{34}+\chi_{34}\Gamma_{12})+\\
&M_{12}\Gamma_{12}+M_{13}\Gamma_{13}+M_{14}\Gamma_{14}+M_{23}\Gamma_{23}+
M_{24}\Gamma_{24}+M_{34}\Gamma_{34}.
\endaligned
$$

%$$
%\aligned \tilde
%H_L&=\left(\frac1{2a}-1\right)M_{12}\left(\frac{M_{13}^2+
%M_{14}^2+
%M_{23}^2+M_{24}^2}2+\Gamma_{12}\right)+\\
%&\left(\frac1{2-2a}-1\right)M_{34}(M_{12}M_{34}+M_{23}M_{14}-M_{13}M_{24}+\Gamma_{34})+\\
%&M_{12}\Gamma_{12}+M_{13}\Gamma_{13}+M_{14}\Gamma_{14}+M_{23}\Gamma_{23}+
%M_{24}\Gamma_{24}+M_{34}\Gamma_{34},
%\endaligned
%$$
%and
%$$
%\aligned  \tilde
%H_{L1}&=\left(2b-1\right)M_{12}\left(\frac{M_{13}^2+ M_{14}^2+
%M_{23}^2+M_{24}^2}2+\Gamma_{12}\right)+\\
%&\left(1-2b\right)M_{34}(M_{12}M_{34}+M_{23}M_{14}-M_{13}M_{24}+\Gamma_{34})+\\
%&M_{12}\Gamma_{12}+M_{13}\Gamma_{13}+M_{14}\Gamma_{14}+M_{23}\Gamma_{23}+
%M_{24}\Gamma_{24}+M_{34}\Gamma_{34}.
%\endaligned
%$$
\endproclaim

The situation with four-dimensional Hess-Appel'rot case is similar
to the three-dimensional case: the  Hamiltonian for the
four-dimensional Hess-Appel'rot system in the first structure is
again a quadratic deformation of $H_{LB}$:
$$
H_{HA}=H_{LB}+J_{13}(-M_{12}M_{23}+M_{14}M_{34})+J_{24}(M_{12}M_{14}-M_{23}M_{34})
$$

Functions $M_{12}$ and $M_{34}$, giving invariant relations for
the four-dimensional Hess-Appel'rot system, are also Casimir
functions for the second Poisson structure (39).

Putting $\chi_{34}=0$, and assuming  $\chi_{12}=1$ in (39) and in
expression for $H_{LB}$, we get the bihamiltonian structure for
the four-dimensional Lagrange top introduced by Ratiu in [37].

In general, in arbitrary  dimension $n$, the standard Poisson structure
 on $so(n)\times so(n)$ is given by:
$$
\{M_{ij}, M_{jk}\}_1=-M_{ik},\quad \{M_{ij},
\Gamma_{jk}\}_1=-\Gamma_{ik},\quad
\{\Gamma_{ij},\Gamma_{kl}\}_1=0,\,\,i,j,k=1,...,n.
$$
In the metric $\Omega=JM+MJ$, the $n$-dimensional Lagrange top is defined
with a Hamiltonian
$$
H_{L}=\frac12\left(2J_1M_{12}^2+(J_1+J_3)\sum_{p=3}^{n}(M_{1p}^2+M_{2p}^2)+2J_3\sum_{3\leq
p<q\leq n}M_{pq}^2\right)+\chi_{12}\Gamma_{12}
$$
  The number of nontrivial integrals of the motion is
$\frac{n(n-1)}{2}-\left[\frac n2\right]$. Casimir functions are
given by (see[37])
$$
tr(\Gamma^{2k}),\quad tr(M\Gamma^{2k+1}). \tag{40}
$$

We use the notation $J_{1}=a,\ J_{3}=1-a,\ \chi_{12}=1$. Let us
introduce a new Poisson structure:
$$
\{\Gamma_{ij},\Gamma_{jk}\}_2=-\Gamma_{ik},\,\{M_{ij},
M_{kl}\}_2=0\,\,\{M_{1l}, M_{2l}\}_2=-1,\,\,l=3,...,n. \tag{41}
$$

The dimension of a symplectic leaf in this structure is
$$
\frac{(n-2)(n-3)}2 -\left[\frac{n-2}2\right]+4(n-2),
$$
hence there are $\frac{n^2-5n+8}2+\left[\frac{n}2\right]$ Casimir
functions:
$$
M_{12}, M_{pq}, Tr(\Gamma^{2k}), \ \ 2<p<q\le n, \ \ \
k=1,...,\left[\frac{n}2\right].
$$

\proclaim{Proposition 12} The $n$-dimensional Lagrange top is a
Hamiltonian system in the Poisson structure {\rm(41)}, compatible
with the standard one. Its Hamiltonian is:
$$
\aligned \tilde
H_{L}&=(2a-1)M_{12}\left(\frac12\sum_{p=3}^n(M_{1p}^2+M_{2p}^2)+\Gamma_{12}\right)+\\
&(1-2a)\sum_{3\leq p<q\leq
n}M_{pq}(M_{1q}M_{2p}-M_{2q}M_{1p}+\Gamma_{pq})+\sum_{1\leq
p<q\leq n}M_{pq}\Gamma_{pq}
\endaligned
$$
\endproclaim
Similarly as in dimension 3 and 4, Hamiltonian for the
Hess-Appel'rot system in arbitrary dimension $n$ is a quadratic
deformation of the Hamiltonian for the $n$-dimensional Lagrange
top:
$$
H_{HA}=H_{L}+\sum_{k=1}^n(J_{13}M_{1k}M_{3k}+J_{24}M_{2k}M_{4k}),
$$
and functions $M_{12}, M_{pq}, p,q\geq 3$, which give the
invariant relations (11),  are Casimir functions for the  Poisson
structure (41).

We can summarize the discussion of this section by saying that
constructed Hess-Appel'rot systems satisfy the following.
\medskip

{\bf (BP) (bi-Poisson condition)}
 {\it There exist a pair of compatible Poisson
structures, such that the system is Hamiltonian with respect to
the first structure, having the Hamiltonian of the form {\rm
(HP)}, such that $f_i$ are Casimir functions with respect to the
second structure.}
\medskip

The invariant relations define symplectic leaves with respect to
the second structure, and the system is Hamiltonian with respect
the first one.

\

\centerline{\bf 11. Back to Kowalevski properties}

\

As we mentioned in the introduction, from the first years of its
history, the Hess-Appel'rot systems were closely related to
Kowalevski's analysis. Investigating the systems constructed in
the first part of this paper, we have noticed that they are
certain perturbations of the form (HP)of integrable systems,
$n$-dimensional Lagrange tops and the Lagrange bitop. We also
observed that perturbing functions $f_i$, which give the invariant
relations, are Casimir functions of the second Poisson structure
and the integrable systems are bihamiltonian corresponding to that
structure. Up to now there is no restriction on the choice of
perturbing functions $b_i$ in (HP). In order to define more
precisely a class of systems which has the same typical dynamical
and analytical properties as the classical and the $n$-dimensional
Hess-Appel'rot case we need to study them in more details.
Finally, after that we will be able to extract the basic ones
leading to the constrains on the functions $b_i$. The correctness
of our choice is illustrated by the Theorem 5 in Section 12. Using
the axioms one can easily construct large number of new examples of
systems of Hess-Appel'rot type (beside semidirect product
$so(n)\times so(n)$ in a study of generalized rigid body systems,
one can consider, for example, semidirect product of $R^n$ and $so(n)$).

To get the right choice of axioms, we have to turn back to the Kowalevski
analysis. First, we are going to introduce some general notions,
see [29].

Suppose a system of ODEs of the form
$$
\dot z_i=f_i(z_1,\dots,z_n), \quad i=1,\dots, n, \tag {42}
$$
is given and  there exist positive integers $g_i, \quad i=1,\dots,
n$, such that
$$
f_i(a^{g_1}z_1,\dots,a^{g_n}z_n)=a^{g_i+1}f_i(z_1,\dots, z_n), \quad i=1,\dots, n.
$$
Then the system (42) is {\it quasi-homogeneous} and numbers $g_i$
are {\it exponents of quasi-homogeneity}. Then, for any complex
solution $C=(C_1,\dots,C_n)$ of the system of algebraic equations:
$$
-g_iC_i=f_i(C_1,\dots,C_n), \quad i=1,\dots, n, \tag {43}
$$
one can define {\it the Kowalevski matrix} $K=K(C)=[K^i_j(C)]$:
$$
K^i_j(C)=\frac{\partial f_i}{\partial z_j}(C)+g_i\delta^i_j.
$$
Eigen-values of the Kowalevski matrix are called {\it the
Kowalevski exponents}. This terminology was introduced in [45]. In
last twenty years, heuristic and theoretical methods in
application of Kowalevski matrix and Kowalevski exponents in study
of integrability and nonintegrability have been actively
developing, see for example [2, 3, 46, 47, 25, 29]. But the notion of
Kowalevski matrix and Kowalevski exponents were introduced by
Kowalevski herself in [28]. The criterion she used [28, p. 183, l.
15-22] to detect a system which is now known as the Kowalevski
top, can be formulated in Yoshida terminology as:

{\bf Kowalevski condition (Kc).} {\it The $6\times 6$ Kowalevski
matrix should have} five different positive integer {\it
Kowalevski exponents}.

Now we return to the study of Hess-Appel'rot systems. The systems
we have constructed {\it are quasi-homogeneous}. Exponents of each
$M$ variable are $g=1$, and for any $\Gamma$ they are equal to
two. We are going now to calculate Kowalevski exponents for the
Hess-Appel'rot systems.
\medskip

{\bf Three-dimensional Hess-Appel'rot case.} Let us denote $(M_1,
M_2, M_3, \Gamma_1, \Gamma_2$, $ \Gamma_3)$ by $(z_1,\dots, z_6)$.
Then the Euler-Poisson equations take the form (42) with
$$
\aligned
f_1&=(J_3-J_1)z_2z_3+J_{13}z_1z_2+z_5;\\
f_2&=(J_3-J_1)z_1z_3+J_{13}(z_3^2-z_1^2)-z_4;\\
f_3&=J_{13}z_2z_3;\\
f_4&=J_3z_5z_3-J_1z_2z_6+J_{13}z_1z_5;\\
f_5&=-J_3z_3z_4+J_1z_1z_6+J_{13}(z_3z_6-z_1z_4);\\
f_6&=J_3z_2z_4-J_1z_1z_5-J_{13}z_3z_5;\\
\endaligned
$$
and $g_i=1, \quad i=1, 2, 3$ and $g_i= 2,\quad i= 4, 5, 6.$ The
invariant relation corresponds to the constraint $c_3=0$. So, we
are looking for solutions $(c_1,c_2,0,c_4,c_5,c_6)$ of the system
of the form (43). One can easily get $c_4=-J_{13}c_1^2+c_2$,
$c_5=-c_1(1+J_{13}c_2)$, $c_6=-(c_1^2+c_2^2)/2$. Then, for $c_1\ne
0$, we get four possible solutions for $(c_1, c_2)$ divided into
two pairs: $(\pm i/J_{13},-1/J_{13})$ and $(\pm
2i/J_{13},-2/J_{13})$. The Kowalevski exponents are
$$
(-1, -2, 2, 4, 3, 3),\qquad (-1, 1, 3, 2, 2, 2),
$$
respectively.
\medskip

Thus, it can easily be seen that classical Hess-Appel'rot system
doesn't satisfy exactly the Kowalevski condition (Kc), although it
is quite close to.
\medskip

{\bf Four-dimensional Hess-Appel'rot systems.} In the
four-dimensional case, there are 12 variables $(M_{12}, M_{13},
M_{14}, M_{23}, M_{24}, M_{34}, \Gamma_{12}, \Gamma_{13},
\Gamma_{14}, \Gamma_{23}, \Gamma_{24},\Gamma_{34})$. Denote them
by $(z_1,\dots, z_{12})$ and
$$
f_i=f_i(z_1,\dots, z_{12}), \quad i=1,\dots,12,
$$
the corresponding right sides of the Euler-Poisson equations.
Exponents of quasi-homogeneity are
$$
g_i=1, \quad i=1,\dots, 6, \qquad g_i=2, \quad i=7,\dots, 12.
$$
We are looking for solutions $(d_1,\dots, d_{12})$ of a system of
the form (43). The invariant relations correspond to constraints
$$
d_1=d_6=0.
$$
From the relations on $f_2, f_3, f_4, f_5$ one can express $(d_8,
d_9, d_{10}, d_{11})$ as functions of $(d_2,\dots, d_5)$. Then,
from the equation on $f_7, f_{12}$ one gets $d_7, d_{12}$ as
functions of $(d_2,\dots, d_5)$. Consider two possible cases
separately.
\medskip

{\bf Example 1.} First consider the case where one of perturbing
constants is equal to zero, say $J_{13}=0$. The solution of the
last system of four equations on four unknowns $(d_2,\dots, d_5)$,
leads to the solution:
$$
d_2=-d_5(=d_9=d_{10}),\quad
d_3=d_4=\sqrt{-1-d_2^2}(=d_{11}=-d_8),\quad (d_7=d_{12}=1),
$$
with arbitrary $d_2$. Computing the Kowalevski matrix, we get
finally the Kowalevski exponents
$$
\aligned
\Bigl (0,-1,3,4,2,1,2,&1,2+2\sqrt{J_{24}^2(1+d_2^2)},2-2\sqrt{J_{24}^2(1+d_2^2)},\\
&1+2\sqrt{J_{24}^2(1+d_2^2)}, 1-2\sqrt{J_{24}^2(1+d_2^2)}\Bigr ).\\
\endaligned
$$

By analyzing correspondent eigen-vectors, we see that eight of
them are tangent to the symplectic leaf, and other four are
transversal to the leaf. Nonintegral Kowalevski exponents
$$
\bigl (2+2\sqrt{J_{24}^2(1+d_2^2)},2-2\sqrt{J_{24}^2(1+d_2^2)},
1+2\sqrt{J_{24}(1+d_2^2)},
1-2\sqrt{J_{24}(1+d_2^2)}\bigr )
$$
correspond to a half of tangential eigen-vectors.
\medskip

{\bf Example 2.} Now, suppose that both $J_{13}$ and $J_{24}$ are
nonzero. To simplify the computations, assume $\chi_{12}=1$,
$\chi_{34}=2$. If $d_2=0$, then there are three cases of
nontrivial solutions of the system (43): \item {1)}
$d_2=0,\;d_3=0,\; d_4=0,\; d_5=\pm \frac i2$; \item {2)}
$d_2=0,\;d_3=\mp \frac i4,\; d_4=\pm i/4,\; d_5=0$; \item {3)}
$d_2=0,\;d_3=\pm \frac i4,\; d_4=\pm \frac i4,\;d_5=0$.

Let us calculate Kowalevski exponents in the last case with
$d_3=i/4, d_4=i/4$. We get first
$$
d_7=\frac{(J_1+J_3)}{48}, d_8=-\frac i2, d_9=0, d_{10}=0, d_{11}=\frac i2,
d_{12}=\frac{(J_1+J_3)}{48},
$$
and then the Kowalevski exponents
$$
\multline
(0,\;-1,\;3,\;4,\;1+\frac{(J_{13}-J_{24})}2,\;1-\frac{(J_{13}-J_{24})}2,\;
\\
2+\frac{(J_{13}-J_{24})}2,\;2-\frac{(J_{13}-J_{24})}2,\;2,\;1,\;2,\;1
).\endmultline
$$
Suppose now that $d_2$ is arbitrary. Then there are two sets of
solutions of (43) of the form
\item {4)}
$$d_2 =s, \;d_3=\mp i\frac {\sqrt{1+s^2(J_1+J_3)^2}}{J_1+J_3},\;d_4=
\pm i\frac {\sqrt{1+s^2(J_1+J_3)^2}}{J_1+J_3},\;d_5=s;
$$
\item {5)}
$$d_2 =s, \;d_3=\pm i\frac {\sqrt{1+s^2(J_1+J_3)^2}}{J_1+J_3},\;d_4=
\pm i\frac {\sqrt{1+s^2(J_1+J_3)^2}}{J_1+J_3},\;d_5=-s.
$$

In the case 4) we get further
$$
\aligned
d_7&=-\frac 1{J_1+J_3},\;\;d_8=i\frac {\sqrt{1+s^2(J_1+J_3)^2}}{J_1+J_3},\;\;d_9=s,\;\\
d_{10}&=-s,\;\;d_{11}=i\frac {\sqrt{1+s^2(J_1+J_3)^2}}{J_1+J_3},\;\;d_{12}=-\frac 1{J_1+J_3}.\\
\endaligned
$$
The Kowalevski exponents are
$$
(0,\;-1,\;3,\;4,\;2,\;2,\;1,\;1,\;1+A,\;1-A,\;2+A,\;2-A)
$$
where
$$
A=2(J_{13}+J_{24})\sqrt{1+s^2(J_1+J_3)^2}.
$$
\medskip

{\bf Five-dimensional Hess-Appel'rot systems.} In this case, there
are 20 variables $(M_{12}, M_{13}, M_{14}, M_{15}, M_{23},
M_{24},M_{25}, M_{34}, M_{35}, M_{45}, \Gamma_{12}, \Gamma_{13},
\Gamma_{14},  \Gamma_{15},\Gamma_{23}, \Gamma_{24},$ $
\Gamma_{25},\Gamma_{34},  \Gamma_{35}, \Gamma_{45})$. As before,
we  denote them  by $(z_1,\dots, z_{20})$. Denoting also by
$$
f_i=f_i(z_1,\dots, z_{20}), \quad i=1,\dots,20,
$$
the corresponding right sides of the Euler-Poisson equations.
Exponents of quasi-homogeneity are
$$
g_i=1, \quad i=1,\dots, 10, \qquad g_i=2, \quad i=11,\dots, 20.
$$
We are looking for solutions $(d_1,\dots, d_{20})$ of a system of
the form (43). The invariant relations correspond to constraints
$$
d_1=d_8=d_9=d_{10}=0.
$$
From the relations on $f_2- f_7$ one can express $(d_{12},...,
d_{17})$ as functions of $(d_2,..., d_7)$. Then, from $f_{11}$ one
gets $d_{11}$ as a function of $(d_2,\dots, d_7)$, and after that,
from $(f_{18},f_{19},f_{20})$ one gets $d_{18}, d_{19}, d_{20}$ as
functions of $(d_2,\dots, d_7)$. The final step is solution of the
system $(f_{12},\dots, f_{17})$ with the unknowns $(d_2,\dots,
d_7)$.
\medskip

{\bf Example 3.} We describe nonzero solutions of (43) under the
assumption $d_5=d_7=0,\;\chi_{12}=1$. There are eight sets of
solutions: \item {1)}
$$d_2=0,\; d_3=0,\; d_4=0,\; d_6=\pm \frac {2i}{J_1+J_3};
$$
\item {2)}
$$
d_2=\pm \frac {2i}{J_1+J_3},\; d_3=0,\; d_4=0,\; d_6=0;
$$
\item {3)}
$$
d_2=\pm \frac i {J_1+J_3},\; d_3=0,\; d_4=0,\; d_6=\pm \frac{i}{J_1+J_3};
$$
\item {4)}
$$
d_2=\pm \frac {2 i}{J_{13}},\; d_3=0,\; d_4=0,\; d_6=- \frac {2}{J_{13}};
$$
\item {5)}
$$
d_2=-\frac {1}{J_{13}},\; d_3=0,\;  d_4=\pm \frac i{J_{13}},\; d_6=0;
$$
\item {6)}
$$
d_2=-\frac {4}{J_{13}},\; d_3=0,\; d_4=\pm \frac {4i}{J_{13}},\;
 d_6=\pm \frac {i}{J_{1}+J_{3}};
$$
\item {7)}
$$
d_2= \frac 1{J_{13}},\; d_3=0, \;d_4=\pm \frac i{J_{13}},\; d_6=\pm \frac {2i}{J_{1}+J_{3}};
$$
\item {8)}
$$
d_2=-\frac 3{J_{13}},\; d_3=\frac {i}{2J_{24}},\; d_4=\pm \frac{\sqrt{J_{13}^2-9J_{24}^2}}{2J_{13}J_{24}},\; d_6=0.
$$

We calculate the Kowalevski exponents in three representative
cases: 1), 4) and 8).

In the case 1) only nonzero $d_i, \;i>15$ are
$d_{16}=2/(J_1+J_3)$, $d_{18}=-2i/(J_1+J_3)$, and (under the
assumption $J_3=1,\;J_{13}=10$) the Kowalevski exponents are
$$
\Bigl(1+A,
\;1-A,\;A,\;-A,\;B,\;1-B,\;1,\;-1,\;-1,\;-1,\;3,\;3,\;3,\;2,\;2,\;2,\;2,\;4,\;4,\;4\Bigr);
$$
where $A=2\frac{\sqrt{100-J_{24}^2}}{J_1+J_3}$ and
$B=2i\frac{J_{24}}{J_1+J_3}$.

In the case 2) $d_{11}=2/(J_1+J_3)$ and $d_{15}=2i/(J_1+J_3)$ are
the only nonzero $d_i,\;i\ge 11$ and the Kowalevski exponents are
$$
(-1,\;-1,\,-1,\;1,\;2,\;2,\;2,\;2,\;3,\;3,\;3,\;4,\;4,\;4,\;
1+A,\;1-A,\;A,\;-A,\;B,\;1-B),
$$
where $A=\frac{4i\sqrt {21}}{J_3+1}$, $B=\frac{\sqrt{20i}}{J_3+1}$ and $J_1=1$,$J_{13}=10$,$J_{24}=4$.

In the case 4) we get that all $d_i, \;i\ge 11$ are zero except
$$
d_{15}=\frac 2{J_{13}},\quad d_{17}=-\frac {2i}{J_{13}}.
$$
The Kowalevski exponents are
$$
(-2,\;0,\;0,\;4,\;-1,\;4,\;-1,\;4,\;-1,\;1,\;1,\;1,\;2,\;2,\;2,\;2,\;3,\;3,\;3,\;3).
$$
In the case 5) the Kowalevski exponents are
$$
(-1,\;0,\,0,\;1,\;1,\;1,\;1,\;1,\;1,\;2,\;2,\;2,\;
2,\;2,\;2,\;2,\;2,\;3,\;3,\;3).
$$
In the case 8) we have the following nonzero $d_i, \;i\ge 11$:
$$
\aligned
d_{12}&=\pm \frac {3i}{4J_{13}},\;d_{13}=\frac 1{4J_{24}}, \; d_{15}=\mp \frac {i \sqrt{J_{13}^2-9J_{24}^2}}{4J_{13}J_{24}},\\
 d_{16}&=\frac 3{4J_{13}},\;
d_{17}=\mp\frac i{4J_{24}},\; d_{18}=\mp \frac { \sqrt{J_{13}^2-9J_{24}^2}}{4J_{13}J_{24}}.
\endaligned
$$
Corresponding Kowalevski exponents are
$$
\bigl(0,\;-1,\;\frac32,\;4,\;\frac72,\;\sqrt {2},\; -\sqrt {2},\; 1+\sqrt {2},\; 1-\sqrt {2},\;1,\;\frac 12,\; 3,\; \frac52,\;1,\; \frac 12,\; 3,\;\frac 52,\;2,\;2,\;2\bigr).
$$
\medskip

{\bf Six-dimensional Hess-Appel'rot systems.} In this case, there
are 30 variables $(M_{12},\dots, M_{56}, \Gamma_{12},\dots,
\Gamma_{56})$. Denoting them  by $(z_1,\dots, z_{30})$, by
$$
f_i=f_i(z_1,\dots, z_{30}), \quad i=1,\dots,30,
$$
the corresponding right sides of the Euler-Poisson equations and
exponents of quasi-homogeneity by
$$
g_i=1, \quad i=1,\dots, 15, \qquad g_i=2, \quad i=16,\dots, 30,
$$
we search to solutions $(d_1,\dots, d_{30})$ of a system of the
form (43). The invariant relations correspond to constraints
$$
d_1=d_{10}=d_{11}=d_{12}=d_{13}=d_{14}=d_{15}=0.
$$
The solution of the system follows the same lines as in the
five-dimensional case.
\medskip

{\bf Example 4.} Under the following assumptions
$$
J_1=1,\;J_3=3,\;\chi_{12}=1,\;d_6=\dots=d_9=0,
$$
we get six sets of solutions of the system (43):
\item {1)}
$$
d_2=0,\;d_3=\frac i2,\;d_4=0,\;d_5=0;
$$
\item{2)}
$$
d_2=-\frac 3{2J_{13}},\;d_3=\frac i{2J_{24}},\;d_4=0,\;
d_5=\sqrt{\frac{J_{13}^2-9J_{24}^2}{J_{13}J_{24}}};
$$
\item{3)}
$$
d_2=-\frac 2{J_{13}},\;d_3=0,\;d_4=i\sqrt{\frac{s^2J_{13}^2+4}{J_{13}}},\;d_5=s;
$$
\item{4)}
$$
d_2=-\frac 1{J_{13}},\;d_3=0,\;d_4=i\sqrt{\frac{s^2J_{13}^2+1}{J_{13}}},\;
d_5=s;
$$
\item{5)}
$$
d_2=-\frac 3{2J_{13}},\;d_3=\frac i{2J_{24}},\;d_4=\sqrt{\frac{J_{13}^2-9J_{24}^2}{J_{13}J_{24}}},\;d_5=0;
$$
\item{6)}
$$
d_2=-\frac 3{2J_{13}},\;d_3=\frac i{2J_{24}},\;d_4=\sqrt{\frac{J_{13}^2-9J_{24}^2-4s^2J_{24}^2J_{13}^2}{J_{13}J_{24}}},\;d_5=s;
$$
where $s$ is an arbitrary parameter.

In the case 1) the only nonzero $d$ are $d_{16}=\frac 12$ and $
d_{21}=\frac i2$. The Kowalevski exponents are
$$
\multline(-1,\;-1,\,-1,\;-1,\;1,\;1,\;1,\;2,\;2,\;2,\;2,\;2,\;2,\;2,\;3,\;3,\;3,\;3,\;\\4,\;4,\;4,\;4,\;
1+A,\;1-A,\;A,\;-A,\;1-iB,\;1-iB,\;iB,\;iB),
\endmultline
$$
where $A=\frac {\sqrt{16-J_{13}^2}}{2}$, $B=J_{13}/2$,
$J_1=1$,$J_3=3$, $J_{24}=4$.

Thus, using into account properties of Kowalevski exponents of
algebraically-integrable Hamiltonian systems, we can conclude that
for the systems we have constructed, functions $b_i$ in the
perturbation formula (HP) should satisfy two conditions:
\medskip

{\bf (QH) (quasi-homogeneity)} {\it The obtained system of
Hamiltonian equations has to be quasi-homogeneous.}
\medskip

In such a case, a Kowalevski matrix exists and we come to the last
condition. Suppose the invariant relations correspond to equations
$z_1=0,\dots, z_k=0$.

Denote by $p$ number of Casimirs: $n=p+2m$, where $2m$ is the
dimension of a general symplectic leaf.

\medskip

 {\bf (ArA) (Arithmetic axiom)} {\it For any nonzero
solution $C=(0,..., 0, c_{k+1},..., c_n)$ of the system {\rm(43)},
the Kowalevski matrix $K(C)$ has $n-p$  eigen-vectors tangent to
the symplectic leaf and $p$ transversal to it. Half of the
Kowalevski exponents which correspond to tangential eigen-vectors
and all of transversal ones are rational numbers. Irrational
numbers among the second half of tangential Kowalevski exponents
are divided into pairs such that the differences are integrally
dependent.}

\

\

\centerline{\bf 12. Description of three-dimensional systems of
Hess-Appel'rot type}

\

Now we would like to derive conditions which determine classical
Hess-Appel'rot system among three-dimensional systems of
Hess-Appel'rot type. More precisely, suppose  a system is given by
the Hamiltonian
$$
H_{1}=H_0+JbM_3, \tag {44}
$$
where $H_0$ is the Hamiltonian of the Lagrange top corresponding
to the first Poisson structure, $M_3$ is its integral and a
Casimir for the second structure, $J$ is a nonzero constant and
$b$ is a function, such that the axioms of the systems of
Hess-Appel'rot are satisfied.

Analyzing the system (43) we come to the first, simple but very
important, properties of such functions $b$. Denote by $f$ the
value of the function $b$ at the point $(\hat c_1,\dots,\hat c_6)$
of nonzero solution of (43).
\medskip

\proclaim{Lemma 10} For a nonzero solution $(\hat c_1,...,\hat
c_6)$ of {\rm(43)} and the value $f=b(\hat c_1,...,\hat c_6)$ it
holds \item {{\rm a)}} $\hat c_1^2+\hat c_2^2=0$ or $f=0$; \item
{{\rm b)}} if $f\ne 0$ and $\hat c_1 =\pm i\hat c_2$ then $f=\mp
i/J$ or $f=\mp 2i/J$.
\endproclaim

The Lemma follows by straightforward calculation. It gives a
possibility to reduce the analysis of functions $b$ to analysis of
their {\bf germs}. By a {\it germ} of a function $b$, we mean $(f,
f_1,\dots, f_6)$, where $f_i$ is the value of $b_{z_i}$ calculated
at points of solutions of system (43).

Lemma 10 leads to important simplifications in a study of
Kowalevski matrices and their characteristic polynomials. We will
denote by $K_1,\dots, K_4$ Kowalevski matrices evaluated on germs,
where
$$
\aligned
K_1&: (f=-i/J, \hat c_1=i \hat c_2),\\
K_2&: (f=2i/J, \hat c_1=-i \hat c_2),\\
K_3&: (f=i/J, \hat c_1=-i \hat c_2),\\
K_4&: (f=-2i/J, \hat c_1=i \hat c_2),\\
\endaligned
$$
and by $\Pch_i,\quad i=1,\dots, 4$ the corresponding
characteristic polynomials, $\Pch_i(w)=\det(K_i-wId)$.

\medskip

\proclaim{Proposition 13} \item {{\rm a)}} Characteristic
polynomials $\Pch_i$ have integer-valued coefficients. \item {{\rm
b)}} These coefficients are $J$-independent. \item {{\rm c)}} The
characteristic polynomial $\Pch_1$ is of the form
$$
\Pch_1(w)=w^6 + A_{15}w^5 + A_{14}w^4+\dots+A_{10}, \tag {45}
$$
where
$$
A_{15}=(-9-2Jf_1\hat c_2), A_{10}=12if_1J\hat c_2(-if_1J\hat c_2
+f_1f_2J^2\hat c_2^2 - iJ^2\hat c_2^2f_2^2 -i). \tag {46}
$$
\endproclaim

The proof of Proposition 13 follows from Arithmetic axiom, Lemma
10 and straightforward calculations.

\proclaim{Proposition 14} For $i=1,\dots,4$ the following relation
holds
$$
\Pch_i(-1)=0.
$$
\endproclaim

Proposition 14 is a well-known property of Kowalevski matrices for
authonomous systems, see [29]. For $i=1$, using the notation
$$
X:=J\hat c_2f_2,\qquad Y:=J\hat c_2f_1,
$$
 we get the following
\medskip
\proclaim {Corollary } For the first germ, the following relation
holds
$$
(2XY-3iY+X-3i)(-Y+X-2)=0. \tag {47}
$$
\endproclaim

In the same notations, from Proposition 13 we get

\proclaim{Lemma 11} If $Y\ne 0$, then
$$
(X-i)(Y-i(X+i))=0. \tag {48}
$$
\endproclaim

By systematical analysis of equations (45-48) finally we come to
the following
\medskip

\proclaim{Theorem 5} The  only non-zero polynomials $b$ which give
systems of Hess-Appel'rot type by relation {\rm(44)} are of the
form
$$
b(z_1,\dots, z_6)=z_1 + kz_3.
$$
All systems of Hess-Appel'rot type of the form {\rm(44)} are the
classical Hess-Appel'rot systems.
\endproclaim
\medskip

{\bf Example.} One of possible solutions of the system (46, 47) is
$X=i,\;Y=-3$. It leads to the function
$$
b(z_1,\dots, z_6)=-3z_1+ iz_2. \tag {49}
$$
Corresponding characteristic polynomial of the Kowalevski matrix
$K_1$ is
$$
\Pch_1(w)=w(w-1)(w-2)(w-3)(w+1)(w+2).
$$
However, the characteristic polynomial of the Kowalevski matrix
$K_3$ is of the form
$$
\Pch_3(w)=(w-1)(w-2)(w-3)(w+1)(2w^2-2w+9).
$$
Thus, the function $b$ given by (49) only partially satisfies the
Arithmetic axiom. \

\

{\bf Acknowledgment. }  The research of both authors was partially
supported by the Serbian Ministry of Science and Technology,
Project Geometry and Topology of Manifolds and Integrable Dynamical Systems.
One of the authors (V. D.) has a pleasure to
thank Professor B. Dubrovin  for helpful remarks; his research was
partially supported by SISSA (Trieste, Italy). The authors would also
like to thank the referee for helpful remarks which improved the
manuscript and for indicating the reference [3].

\

\

\centerline {\bf References}

\

\item {1} Adler, M.,van Moerbeke, P.: Linearization of
Hamiltonian Systems, Jacobi Varieties and Representation Theory.
Advances in Math.{\bf 38}, 318-379 (1980)

\item {2} Adler, M., van Moerbeke, P.: The complex geometry of the
Kowalewski -Painlev\' e analysis. Invent. Math. {\bf 97}, 3-51 (1989)

\item {3} Adler, M., van Moerbeke, P., Vanhaeke, P.: {\it Algebraic
integrability,  Painlev\' e geometry and Lie algebras}, Springer-Verlag, Berlin,
2004

\item{4} Appel'rot, G.G.: The problem of motion of a rigid body
about a fixed point.  Uchenye Zap. Mosk. Univ. Otdel. Fiz.
Mat. Nauk, No. 11 , 1-112 (1894)

\item{5} Arbarello, E., Cornalba, M., Griffiths, P.A., Haris J.:
{\it Geometry of algebraic curves}. Springer-Verlag, 1985

\item{6} Arnol'd, V.I.: {\it Mathematical methods of classical
mechanics}. Moscow: Nauka, 1989 [in Russian, 3-rd edition]

\item{7} Arnol'd, V.I., Kozlov, V.V., Neishtadt, A.I.: {\it
Mathematical aspects of classical and celestial mechanics/ in
Dynamical systems III}. Berlin: Springer-Verlag, 1988

\item {8} Audin, M.: {\it Spinning Tops}. Cambridge studies in
Advanced Mathematics 51, 1996

\item{9} Beauville, A.: Prym varieties and Schottky problem.
Inventiones Math. {\bf 41},  149-196 (1977)

\item{10} Belokolos, E.D., Bobenko, A.I., Enol'skii, V.Z., Its, A.R.,
 Matveev, V.B.: {\it Algebro-geometric approach to
nonlinear integrable equations}, Springer series in Nonlinear
dynamics, 1994

\item {11} Bobenko, A.I., Reyman, A.G., Semenov-Tien-Shansky, M.A.:
The Kowalewski top 99 years later: a Lax
pair, generalizations and explicit solutions. Comm. Math.
Phys. {\bf 122}, 321-354 (1989)

\item {12} Bogoyavlensky, O.I.: Integrable Euler equations on Lie
algebras arising in physical problems. Soviet Acad Izvestya,
{\bf 48}, 883-938 (1984) [in Russian]

\item {13} Borisov, A.V., Mamaev, I.S.: {\it Dynamics of rigid body}.
 Moskva-Izhevsk: RHD, 2001, [in Russian]

\item {14} Dalalyan, S.G.: Prym varieties of unramified double
coverings of the hyperelliptic curves. Uspekhi Math. Naukh
{\bf 29}, 165-166 (1974), [in Russian]

\item {15} Dragovi\' c, V., Gaji\' c, B.: An L-A pair for the
Hess-Apel'rot system and a new integrable case for the
Euler-Poisson equations on $so(4) \times so(4)$. Roy. Soc. of
Edinburgh: Proc A {\bf 131}, 845-855 (2001)

\item {16} Dragovi\' c, V., Gaji\' c, B.: The Lagrange bitop on
$so(4)\times so(4)$ and geometry of Prym varieties. American
Journal of Mathematics, {\bf 126}, 981-1004, (2004)
%math-ph/0201036, Preprint SISSA, 79/FM 2001

\item {17} Dubrovin, B.A.: Completely integrable Hamiltonian
systems connected with matrix operators and Abelian varieties.
 Func. Anal. and its Appl. {\bf 11}, 28-41 (1977) [in
Russian]

\item {18} Dubrovin, B.A.: Theta-functions and nonlinear
equations. Uspekhi Math. Nauk {\bf 36}, 11-80 (1981) [in
Russian]

\item {19} Dubrovin, B.A., Krichever, I.M., Novikov, S.P.:
Integrable systems I. In: {\it Dynamical systems IV}. Berlin:
Springer-Verlag, 1990, pp.173-280

\item {20} Dubrovin, B.A., Matveev, V.B., Novikov, S.P.: Nonlinear
equations of Kortever-de Fries type, finite zone linear operators
and Abelian varieties. Uspekhi Math. Nauk {\bf 31},
55-136 (1976) [in Russian]

\item {21} Fay, J.D.: {\it Theta functions on Riemann surfaces},
Lecture Notes in Mathematics, vol. 352, Springer-Verlag, 1973

\item {22} Gavrilov, L., Zhivkov, A.: The complex geometry of
Lagrange top. L'Ensei\-gnement Math\' ematique {\bf 44}, 133-170 (1998)

\item{23} Gel'fand, I.M., Dorfman, I.Ya.: Hamiltonian operators and
algebraic structures connected whit them. Funct. Anal.
Appl. {\bf 13} (4), 13-30 (1979), [in Russian]

\item{24} Golubev, V.V.: {\it Lectures on integration of the
equations of motion of a rigid body about a fixed point} Moskow:
Gostenhizdat, 1953 [in Russian]; (English translation:
Philadelphia: Coronet Books, 1953).

\item {25} Goriely, A.: Integrability, partial integrability and
nonintegrability for systems of ODE. J. Math. Phys {\bf 37},
1871-1893 (1996)

\item{26} Griffiths, P. A.: Linearizing flows and a cohomological
interpretation of Lax equations. {\it American Journal of Math} {\bf 107}
(1983), 1445-1483.

\item{27} Hess, W.: Ueber die Euler'schen Bewegungsgleichungen und
{\"u}ber eine neue particul{\"a}re L{\"o}sung des Problems der
Bewegung eines starren K{\"o}rpers um einen festen. Punkt.
Math. Ann. {\bf 37}, 178-180 (1890)

\item{28} Kowalevski, S.: Sur le probl\` eme de la rotation d'un
corps solide autour d'un point fixe. Acta Math. {\bf 12},
 177--232 (1889)

\item {29} Kozlov, V.V.: {\it Symmetries, topology, resonansies in
Hamiltonian mechanics}. Izevsk, 1995, p. 429 [in Russian]

\item{30} Krichever, I.M.: Algebro-geometric methods in the theory
of nonlinear equations. Uspekhi Math. Naukh {\bf 32},
183 - 208 (1977), [in Russian]

\item{31} Leimanis, E.: {\it The general problem of the motion of
coupled rigid bodies about a fixed point}. Berlin, Heidelberg,
New York: Springer-Verlag, 1965

\item{32} Manakov, S.V.: Remarks on the integrals of the Euler
equations of the $n$-dimensional heavy top. Funkc. Anal.
Appl. {\bf 10}, 93-94 (1976) [in Russian]

\item {33} van Moerbeke, P., Mumford, D.: The spectrum of
difference operators and algebraic curves. Acta Math. {\bf
143}, 93-154 (1979)

\item {34} Mumford, D.: Theta characteristics of an algebraic
curve. Ann. scient. Ec. Norm. Sup. 4 serie {\bf 4},
181-192 (1971)

\item {35} Mumford, D.: Prym varieties 1.  A collection of
papers dedicated to Lipman Bers, New York: Acad. Press
 325-350 (1974)

\item{36} Nekrasov, P.A.: Analytic investigation of a certain case
of motion of a heavy rigid body about a fixed point. Mat.
Sbornik {\bf 18}, 161-274 (1895)

\item{37} Ratiu, T.: Euler-Poisson equation on Lie algebras and the
N-dimensional heavy rigid body. American Journal of Math.
{\bf 104}, 409-448 (1982)

\item {38} Ratiu, T., van Moerbeke, P.: The Lagrange rigid body
motion. Ann. Ins. Fourier, Grenoble {\bf 32},
211-234 (1982)

\item {39} Reyman, A.G., Semenov-Tian-Shansky, M.A.: Lax
representation with spectral parameter for Kowalevski top and its
generalizations. Funkc. Anal. Appl. {\bf 22}, 87-88 (1982)
[in Russian]

\item {40} Shokurov, V.V.: Algebraic curves and their Jacobians,
In: Algebraic Geometry III, Berlin: Springer-Verlag, 1998, pp.219-261

\item {41} Shokurov, V.V.: Distinguishing Prymians from Jacobians.
Invent. Math. {\bf 65}, 209-219 (1981)

\item{42} Sretenskiy, L.N.: On certain cases of motion of a heavy
rigid body with gyroscope. Vestn. Mosk. Univ. No. 3,
60-71 (1963), [in Russian].

\item{43} Trofimov, V.V., Fomenko, A.T.: {\it Algebra and
geometry of integrable Hamiltonian differential equations }.
Moscow: Faktorial, 1995 [in Russian]

\item {44} Whittaker, E.T.: {\it A treatise on the analytical dynamics
of particles and rigid bodies}. Cambridge at the University Press,
1952, p.456

\item {45} Yoshida, H.: Necessary conditions for the existence of
algebraic first integrals, I: Kowalevski's exponents. J.
Celest. Mech. {\bf 31}, 363-379 (1983)

\item {46} Yoshida, H.: A criterion for the nonexistence of an
additional analytic integral in Hamiltonian systems with $n$
degrees of freedom. Phys. Lett. A {\bf 141}, 108-112 (1989)

\item {47} Yoshida, H., Gramaticos, B., Ramani, A.: Painlev\' e
Resonances versus Kowalevski exponents. Acta Applicanda
Mathematicae {\bf 8},  75-103 (1987)

\item{48}  Zhukovski, N.E.: Geometrische interpretation des
Hess'schen falles der bewegung eines schweren starren korpers um
einen festen Punkt. Jber. Deutschen Math. Verein. {\bf 3}, 62-70 (1894)

\item {49} Ziglin, S.L.: Branching of solutions and nonexistence
of first integrals in Hamiltonian mechanics. Funct. Anal.
Appl. {\bf 16}, 181-189 (1983), [in Russian]

\item {50} Ziglin, S.L.: Branching of solutions and nonexistence
of first integrals in Hamiltonian mechanics II. Funct. Anal.
Appl. {\bf 17}, 6-17 (1984), [in Russian].

\end